\documentclass[a4paper,11pt]{article}
\usepackage{jheppub} 
\usepackage{lineno}

\usepackage[utf8]{inputenc}
\usepackage[normalem]{ulem}
\usepackage{graphicx}
\usepackage{dcolumn}
\usepackage{array} 
\usepackage{bm}
\usepackage{color}
\usepackage[dvipsnames]{xcolor}

\usepackage{url}
\usepackage{xspace}
\usepackage{slashed}
\usepackage{multirow,bigstrut}
\usepackage{mathrsfs} 
\usepackage{relsize,amsmath}
\usepackage[geometry]{ifsym}
\usepackage{amssymb}
\usepackage{pifont}
\usepackage{physics}
\usepackage[compat=1.1.0]{tikz-feynman}
\usepackage{cleveref}

\usepackage{rotating}
\usepackage{ragged2e}

\usepackage{fontawesome} 
\definecolor{blue-violet}{rgb}{0.33, 0.17, 0.89}

\renewcommand{\phi}{\varphi}

\newcounter{CommentCount}
\definecolor{MH}{rgb}{0.0,0.6,9}
\definecolor{palatinate}{rgb}{0.494, 0.192, 0.482}

\definecolor{teal}{HTML}{008080}

\usepackage{siunitx}

\DeclareSIUnit \s {\second}
\DeclareSIUnit \ns {\nano\second}
\DeclareSIUnit \mus {\micro\second}
\DeclareSIUnit \ms {\milli\second}
\DeclareSIUnit \MB {\mega\byte}
\DeclareSIUnit \GB {\giga\byte}
\DeclareSIUnit \TB {\tera\byte}
\DeclareSIUnit \PB {\peta\byte}
\DeclareSIUnit \Mbps {\mega\bit/\s}
\DeclareSIUnit \Gbps {\giga\bit/\s}
\DeclareSIUnit \Tbps {\tera\bit/\s}
\DeclareSIUnit \Pbps {\peta\bit/\s}
\DeclareSIUnit \kton {\kilo\tonne} 
\DeclareSIUnit \kt {\kilo\tonne}
\DeclareSIUnit \Mt {\mega\tonne}
\DeclareSIUnit \eV {\electronvolt}
\DeclareSIUnit \keV {\kilo\electronvolt}
\DeclareSIUnit \MeV {\mega\electronvolt}
\DeclareSIUnit \GeV {\giga\electronvolt}
\DeclareSIUnit \TeV {\tera\electronvolt}
\DeclareSIUnit \PeV {\peta\electronvolt}
\DeclareSIUnit \EeV {\exa\electronvolt}
\DeclareSIUnit \m {\meter}
\DeclareSIUnit \cm {\centi\meter}
\DeclareSIUnit \in {\inchcommand}
\DeclareSIUnit \km {\kilo\meter}
\DeclareSIUnit \kV {\kilo\volt}
\DeclareSIUnit \kW {\kilo\watt}
\DeclareSIUnit \MW {\mega\watt}
\DeclareSIUnit \MHz {\mega\hertz}
\DeclareSIUnit \mrad {\milli\radian}
\DeclareSIUnit \year {years}
\DeclareSIUnit \POT {POT}
\DeclareSIUnit \sig {$\sigma$}
\DeclareSIUnit\parsec{pc}
\DeclareSIUnit\lightyear{ly}
\DeclareSIUnit\foot{ft}
\DeclareSIUnit\ft{ft}
\DeclareSIUnit \ppb{ppb}
\DeclareSIUnit \ppt{ppt}
\DeclareSIUnit \samples{S}
\DeclareSIUnit \pe{PE}
\DeclareSIUnit \T{T}

\newcommand{\enu}{\E_\enu}

\definecolor{myred}{cmyk}{0,1,1,0.55}
\definecolor{mygreen}{rgb}{0.27, 0.64, 0.48}
\definecolor{mygray}{gray}{.95}
\definecolor{blue-violet}{rgb}{0.33, 0.17, 0.89}

\definecolor{myred}{cmyk}{0,1,1,0.55}
\definecolor{mygreen}{rgb}{0.27, 0.64, 0.48}
\definecolor{mygray}{gray}{.95}

\arxivnumber{2602.06111} 

\title{\boldmath \Large Dark Matter Heating of Compact Stars Beyond Capture: \\
        \large \it A Relativistic Framework for Energy Deposition by Particle Beams}

\author[a]{Jaime Hoefken Zink,}

\author[b]{Shihwen Hor,}

\author[c,d]{and Maura E. Ramirez-Quezada}

\affiliation[a]{ National Centre for Nuclear Research, Pasteura 7, Warsaw, PL-02-093, Poland,}
\affiliation[b]{ Tsung-Dao Lee Institute, Shanghai Jiao Tong University, No.~1 Lisuo Road, Shanghai, 201210, China,}
\affiliation[c]{Johannes Gutenberg-Universität Mainz, 55099 Mainz, Germany, and}
\affiliation[d]{Dual CP Institute of High Energy Physics, C.P. 28045, Colima, M\'exico.}

\emailAdd{jaime.hoefkenzink@ncbj.gov.pl}
\emailAdd{shihwen@sjtu.edu.cn}
\emailAdd{mramirez@uni-mainz.de}

\abstract{Compact astrophysical objects, such as neutron stars and white dwarfs, can act as detectors of energetic particle fluxes originating from astrophysical accelerators. While most existing capture and heating calculations assume isotropic very low energetic incident fluxes from the halo dark matter, many realistic sources produce highly directional beams or jets, for which gravitational focusing, trajectory multiplicity, and local energy deposition must be treated consistently. In this work, we develop a general relativistic formalism to compute the local density, capture probability, and energy 
deposition of particles arriving as directed beams onto compact objects. The framework is based on the mapping of an asymptotic particle flux to local densities through geodesic congruences, allowing for gravitational focusing, multi–stream regions, and optical depth effects to be incorporated in a unified way. The formalism applies to arbitrary particle species and interaction models, and separates capture from through–going energy deposition in a frame–consistent manner. As an explicit application, we consider relativistic particle beams generated in astrophysical jets and evaluate their interaction with two compact objects samples: a white dwarf and a neutron star. In particular, we illustrate the framework using boosted dark matter produced in a list of 324 blazars as a representative case study, computing the resulting fluxes and the associated heating in the selected stars. Additional regimes such as the interaction roof and geometric limit are discussed, highlighting the conditions under which compact objects can efficiently convert incident beam energy into observable heating. We particularly show that for fermionic dark matter interacting through an axion-like-particle mediator with suppressed low energy cross sections, boosted dark matter heating is much stronger than the standard halo dark matter heating. We reach sensitivities for direct detection cross sections of dark matter masses of $10$ MeV of the order of $10^{-39} - 10^{-27} \, \mathrm{cm}^2$ depending on the mediator mass by considering the coolest observed white dwarf, WD J2147-4035.}

\begin{document}
\hfill {\tt MITP-26-004}
\maketitle
\flushbottom

\section{Introduction}
\label{sec:Introduction}

Dark matter (DM) remains one of the most fundamental open problems in particle physics and cosmology, as it lies beyond the explanatory power of the Standard Model (SM) despite multiple observational indications of its presence. Considerable experimental effort has been devoted to uncovering this elusive component of the Universe, including the construction of large-scale and costly direct-detection experiments~\cite{Bernabei:2013xsa, Bernabei:2018jrt, Armengaud:2013rta, ZEPLIN-III:2010cnv, Akimov:2011tj, CoGeNT:2012sne, CRESST:2015txj, XENON100:2016sjq, DRIFT:2016utn, Leyton:2016nit, DAMIC:2020cut, DM-Ice:2016snk, CDEX:2014dae, CDEX:2019hzn, XMASS:2018xyo, SuperCDMS:2019jxx, SuperCDMS:2017mbc, KIMS:2018hch, DarkSide:2022knj, LUX:2020yym, PICO:2016kso, PICO:2019vsc, EDELWEISS:2018tde, PandaX:2015gpz, PandaX-II:2020oim, PandaX:2023tfq, NEWS-G:2022kon, CRESST:2022dtl, DEAPCollaboration:2021raj, COSINE-100:2021zqh, Amare:2021yyu, CRESST:2020tlq, CDEX:2022rxz, EDELWEISS:2022ktt, Hamaide:2021hlp, Das:2022srn, SENSEI:2019ibb, XENON:2020rca, XENON:2023cxc, LZ:2022lsv}. Most of these searches are based on the hypothesis that DM interacts with SM particles through non-gravitational forces and are designed to probe the expected local population of cold DM at extremely low energies. To date, however, no definitive signal has been observed. This lack of success has motivated the exploration of alternative and more innovative detection strategies. In this work, we model a mechanism in order to use compact stars as detectors of boosted DM, extending previous works on halo detection to more general flux configurations, especially in the high energy regime, but without restricting the treatment to a specific energy range.

Assuming that DM interacts with SM particles via a dark or hidden force naturally opens up the possibility that it can be accelerated by highly energetic SM particles. Several such boosting mechanisms have been investigated in the literature, including cosmic-ray–boosted DM (CRBDM)~\cite{Bringmann:2018cvk, Dent:2019krz, Lei:2020mii, Super-Kamiokande:2022ncz, Maity:2022exk, Wang:2023wrx, Bell:2023sdq, PandaX:2024pme, Cappiello:2024acu, Ghosh:2024dqw, Diurba:2025lky, Gustafson:2025dff, LZ:2025iaw}, DM boosted through semi-annihilation or decay processes (DABDM)~\cite{Kopp:2015bfa, Necib:2016aez, Fornal:2020npv, Toma:2021vlw, Aoki:2023tlb, BetancourtKamenetskaia:2025noa, Clark:2024baf}, neutrino-induced boosted DM ($\nu$BDM)~\cite{Yin:2018yjn, Jho:2021rmn, Das:2021lcr, Lin:2022dbl, Lin:2023nsm, Lin:2024vzy, DeRomeri:2023ytt, Das:2024ghw, Ghosh:2024dqw, Sun:2025gyj}, and blazar boosted DM (BBDM)~\cite{Wang:2021jic, Granelli_2022ysi, Cline:2022qld, Bhowmick:2022zkj, DeMarchi:2024riu, CDEX:2024qzq, Jeesun:2025gzt, Zapata:2025huq, Wang:2025ztb, Dev:2025czz, DeMarchi:2025uoo, Barillier:2025xct}, among other proposed scenarios. BBDM could in principle produce the strongest high energetic DM fluxes.

Compact astrophysical objects provide unique environments in which energetic particles can interact with dense matter and strong gravitational fields. Because of their high densities, small radii, and long lifetimes, such objects can act as effective detectors of weakly interacting or rare particle fluxes, converting incident kinetic energy into heat through scattering and capture processes. This idea has motivated a broad literature exploring the role of compact stars as probes of particle physics beyond the SM, as well as of high–energy astrophysical phenomena~\cite{Bertone:2007ae,McCullough:2010ai,Hooper:2010es,Amaro-Seoane:2015uny,Panotopoulos:2020kuo,Biswas:2022cyh}.

Among compact objects, neutron stars (NSs) and white dwarfs (WDs) are particularly attractive targets. NSs exhibit extreme conditions of high densities and strong gravitational fields, compared to any other stellar object other than black holes. WDs are abundant in the Galaxy, their internal structure is comparatively well understood. For both kinds of stars, their extreme densities allow incident particles to be accelerated to high velocities as they fall into the stellar gravitational potential. Interactions with stellar matter can lead either to gravitational capture or to energy deposition along the particle trajectory, resulting in heating of the stellar interior. The theoretical foundations of particle capture in stars were originally developed by Press and Spergel and later refined by Gould~\cite{Press:1985ug,Gould:1987ir,Gould:1987ju}, providing the basis for computing capture rates and energy loss in dense astrophysical media. More generally, heating mechanisms in compact stars arising from particle interactions have been explored in a variety of particle–physics scenarios~\cite{Bramante:2015dfa, Gani:2018mey, Garani:2018kkd, Garani:2019fpa, Acevedo:2019gre, Bell:2020jou, Bell:2020lmm, Bell:2021fye, Bell:2023sdq, Davoudiasl:2025gxn}.

Most existing capture and heating calculations assume isotropic and low energetic incident particle populations. However, compact stars may also be exposed to highly directional fluxes originating from energetic astrophysical environments, such as BBDM, and also to high energetic isotropic fluxes, such as CRBDM. In both cases, the standard isotropic-low energetic formalism must be generalized to account for high energy components, anisotropic phase–space distributions, gravitational focusing, and the possibility that multiple particle trajectories intersect the same spatial region inside the star. In addition, sufficiently energetic particles may deposit significant amounts of energy through scattering even if they do not become gravitationally bound, providing an important heating channel that is distinct from capture and must be treated consistently.

A well–motivated example of a directed, relativistic particle flux, as mentioned above, is provided by blazar boosted dark matter, in which DM particles acquire large kinetic energies through interactions with protons in blazars. Such boosted populations have been studied extensively in the context of terrestrial experiments and astrophysical environments. In previous work, we initiated a description of particle capture in white dwarfs across a wide range of kinetic energies using simplified, isotropic flux assumptions~\cite{HoefkenZink:2024hor}. These studies highlight the need for a more general treatment capable of handling realistic, source–motivated and anisotropic particle fluxes.

In this work, we develop a general formalism for computing capture and heating of compact objects by directed particle beams. As we will also show, this formalism is also valid for isotropic fluxes, as long as the compact object is spherically symmetric. Starting from an asymptotic flux defined far from the star, we construct the local particle density using geodesic congruences, allowing gravitational focusing, multi–stream regions, and optical depth effects to be treated consistently. The framework separates gravitational capture from through–going energy deposition and is applicable to arbitrary particle species and interaction models. As an explicit application, we illustrate the formalism using fermionic DM beams generated in astrophysical jets and interacting with them through an axion-like-particle (ALP) mediator, with boosted DM serving as a representative case study rather than the primary focus of the analysis. We want this study to serve as a model-independent paradigm and guide for future analyses and computations, involving DM jets and compact star heating.

In \Cref{sec:Observables}, we briefly discuss the two compact objects treated in the present work: white dwarfs and neutron stars. \Cref{sec:BDM_formalism} is the core of our study: it develops the formalism from general relativity to compute the capture heating rates of compact objects by DM jets. It also analyses extreme cases, such as the interaction roof, when all particles going through the star interact with it, and the geometric limit, when all particles are captured, i.e., gravitationally bound to the star. It also shows how isotropic fluxes can be treated as directed jets, as long as the star is not quickly rotating. \Cref{sec:examples} show the results of computations performed for a concrete model of fermionic DM with an ALP mediator. We also regard a DM flux originated from a list of 324 blazars~\cite{Rodrigues:2023vbv}, whose fitted parameters allow for a full computation of the generated fluxes. Finally, in \Cref{sec:conclusions} we discuss the conclusions of what was obtained and future perspectives. The reader may also refer to the appendices for more details about the computations.

\section{Astrophysical Objects}
\label{sec:Observables}

\begin{table}
\caption{Physical properties of selected compact objects.}
\resizebox{\textwidth}{!}{
\begin{tabular}{cccccccc}
\hline
\textbf{Benchmark} & $T_{\mathrm{eff}}$ [K] & $T_\star$ [K] & Log(g) [cgs] & $M_\star$ [$M_\odot$] & $R_\star$ & More \\
\hline
WD & $3048$ & $3050$ & 8.195 & 0.69 & 1.2 $R_\oplus$ & Spectral Type: DZQH \\
NS & $5.8 - 11.6 \times 10^5$ & $7.3 \times 10^5$ & 14.4 & 1.4 & 12.1 km & Name: RX J1856.5-3754 \\
\hline
\end{tabular}}
\label{tab:co_prop}
\end{table}

We focus on compact stellar remnants, specifically white dwarfs and neutron stars , as astrophysical targets. For all compact objects considered in this work, stellar radial profiles are obtained by solving the Tolman--Oppenheimer--Volkoff (TOV) equations for static, spherically symmetric configurations~\cite{PhysRev.55.364,PhysRev.55.374}, coupled to the appropriate equation of state (EOS). For a given EOS, the TOV system is numerically integrated for a range of central energy densities and interpolated to reproduce the desired stellar mass, yielding radial profiles of density, pressure, enclosed mass, and thermodynamic quantities. Object-specific assumptions regarding composition and the choice of EOS are described in the corresponding subsections below.

\subsection{White Dwarfs in the Milky Way}
\label{sec:WDs_MW}

Most stars in our Galaxy will eventually evolve into white dwarfs. These compact remnants, supported by electron degeneracy pressure, typically have masses around $0.6\,M_\odot$, radii of order of the Earth, and effective temperatures in the range $10^3$--$10^4\,\mathrm{K}$. They are generally composed of carbon, oxygen, or an admixture of both, with outer layers dominated by hydrogen or helium. They exhibit extreme conditions, providing excellent astrophysical laboratories to probe energetic particle interactions.


For the purpose of this work, we focus on the coolest WD ever observed, WD J2147-4035~\cite{Elms:2022oic}, as a benchmark for our capture and energy deposition analysis. The star is selected from the catalogue provided by the Gaia mission~\cite{2021MNRAS.508.3877G}. This star is a DZQH-type WD located within 28~pc from the Sun in the constellation Grus, with a mass of $0.69\,M_\odot$. This mass ensures that the equation of state can be safely approximated as temperature-independent and minimizes uncertainties. Its macroscopic properties are summarized in the first row of \Cref{tab:co_prop}. To validate the stellar parameters, we perform cross-checks with Gaia DR3~\cite{gaia2022vizier} and EDR3~\cite{2021A&A...649A...6G} data products when available, extracting the mass, effective temperature, and surface gravity.

Stellar profiles are constructed following the general procedure described above, adopting a zero-temperature equation of state including Coulomb, Thomas--Fermi, exchange, and correlation corrections~\cite{salpeter61}, and assuming a carbon core composition.

\subsection{Neutron Stars}
\label{sec:NS}

Neutron stars are extremely dense stellar remnants with a stratified structure consisting of an outer crust and an inner core. The crust typically extends to depths of up to a few hundred meters below the surface, while the core reaches densities $\rho \gtrsim \mathcal{O}(10^{14})\,\mathrm{g\,cm^{-3}}$. For the purpose of this work, we assume the standard npe$\mu$ matter, i.e. NSs composed primarily of neutrons, protons, electrons, and muons, and we neglect the presence of hyperons.

To model the NS macroscopic structure we adopt a cold, non-accreting equation of state. Many approaches exist in the literature~\cite{Pearson:2018tkr,Akmal:1998cf,Rikovska-Stone:2006gml,Goriely:2010bm,Pearson:2012hz,PhysRevC.88.024308,Potekhin:2013qqa,Kojo:2014rca,Baym:2017whm,Annala:2019eax}. In this work we focus on old and cold stars and employ the Brussels--Montreal BSk25 EoS~\cite{Xu:2012uw,PhysRevC.88.061302,Perot:2019gwl}. As a benchmark, we adopt an NS mass $M_{\rm NS}=1.4\,M_\odot$, which corresponds to the closest NS: RX J1856.5-3754. The BSk25 EoS yields a radius $R_{\rm NS}=12.1\,\mathrm{km}$ and a central density $\rho_c = 7.45\times 10^{14}\,\mathrm{g\,cm^{-3}}$~\cite{Xu:2012uw,PhysRevC.88.061302,Perot:2019gwl}. These benchmark macroscopic properties are the inputs used throughout our NS estimates and are shown in the second row of \Cref{tab:co_prop}.

\section{Boosted Dark Matter in compact objects}
\label{sec:BDM_formalism}

\subsection{General formalism}
The formalism commonly employed to compute capture rates in stellar objects~\cite{Busoni:2017mhe, Busoni:2022wzc, Bell:2020jou, Bell:2020lmm, Anzuini:2021lnv} has been developed under the assumption of an isotropic population of low energetic DM surrounding the star. In this work, instead, we treat an \emph{incident beam} of particles arriving from far away with a preferred direction. This setup is relevant whenever the incoming population is sourced non-locally (e.g.\ by astrophysical accelerators) and reaches the target as a directed flux rather than as a quasi-static, isotropic distribution. We use BBDM as a concrete case study, but the construction below is very general and applies to any particle species approaching in a fixed direction far from the star, with a trajectory described by geodesic propagation and able to interact with the stellar medium. We will consider interactions in the star, regardless of whether the DM particles are able to escape the star or not. The standard capture rate is recovered as a special case by restricting scatterings to those that leave the outgoing particle gravitationally bound.

One of the earliest treatments of capture in stars was given by Press and Spergel~\cite{Press:1985ug}, and later by Gould~\cite{Gould:1987ir, Gould:1987ju}, among other early studies. The capture rate is understood as the number of particles per time that get trapped in the gravitational potential of the star. In this framework, the differential capture rate can be written as~\cite{Gould:1987ir}
\begin{equation}
\label{eq:capture_rate_original}
dC = \frac{dN_\chi}{dt} \times \frac{dl}{\omega} \times \Omega^-_{v_e}(\omega)\,.
\end{equation}
From this expression, three elements with a clear physical interpretation can be identified:  
(i) $\,dN_\chi/dt$, the rate at which particles enter a given volume element of the star;  
(ii) $\,dl/\omega$, the time spent by each particle in regions where scattering may occur; and  
(iii) $\,\Omega^-_{v_e}(\omega)$, the probability rate of a particle to interact such that a single collision reduces the particle's velocity below the escape velocity, leading to gravitational binding. Here, $dl$ denotes the length of the particle trajectory within a given radial shell and $\omega$ its velocity. The local escape speed is defined as $v_e \equiv \sqrt{1 - g_{tt}(r)}$, where $g_{tt}(r)$ is the time component of the spacetime metric at radius $r$.

Since we can no longer assume a uniform flux through spherical shells, we must adopt a different approach. As a first step, we reinterpret the previous expression in a more convenient language. This reformulation replaces the picture of particles crossing spherical shells isotropically with congruences of geodesics, i.e., families of non-intersecting trajectories parametrized by their impact parameters and weighted by their fluxes. We therefore write the differential capture rate as
\begin{equation}
\label{eq:capture_rate_new}
dC = dn_\chi(T^\infty_\chi, r, \phi)\,
      \Omega^-_{v_e}\!\left(\omega(T^\infty_\chi, r)\right)\,
      dV \,,
\end{equation}
where $dn_\chi$ denotes the number density of particles with kinetic energy in the interval $[T_\chi, T_\chi + dT_\chi]$, evaluated locally at each position inside the star. Here $T^\infty_\chi$ is the kinetic energy of the particle at infinity, and $\omega$ is the local velocity inside the star, which depends on both $T^\infty_\chi$ and the radial position where the interaction takes place. In what follows, $dC$ is used as a compact notation for the differential \emph{beam-induced} rate element. Since particles with different asymptotic energies follow distinct geodesics, the corresponding number densities must be computed separately for each interval $[T_\chi, T_\chi + dT_\chi]$.

To obtain these densities, we consider the system at infinity, where the incident flux originates. Let us assume the flux arrives from infinity propagating along the $\hat{x}$ direction. We construct a volume containing a number of particles $dN$ corresponding to an energy differential bin $[T_\chi^\infty, T_\chi^\infty + dT_\chi^\infty]$ of the flux $d\Phi_\chi / dT_\chi^\infty$. We take $\lambda \equiv \tau$ to be the affine parameter of the geodesic equations, equal to the particle's proper time.

Our goal is to track how an initially uniform beam is distorted by gravitational focusing, translating the flux at infinity into a spatial density inside the star. With these definitions, we obtain
\begin{equation}
\label{eq:dN}
dN 
= dT_\chi^\infty \, \frac{d\Phi_\chi}{dT_\chi^\infty} \, dt \, dA_{\perp}
= dT_\chi^\infty \, \frac{d\Phi_\chi}{dT_\chi^\infty} \, dV_0 \,,
\end{equation}
where $dA_{\perp}$ is the area perpendicular to the incident flux and $dV_0$ is the infinitesimal volume swept by the particles during a coordinate-time interval $dt$.

The particle density at each point inside the star is obtained by evolving the initial volume element $dV_0$ along the geodesic congruence and dividing the number of particles by the corresponding volume at position $\vec r$, denoted $dV_r$,
\begin{equation}
\label{eq:dV_0}
dn_\chi
= \frac{dN}{dV}
= dT_\chi^\infty \, \frac{d\Phi_\chi}{dT_\chi^\infty}
\frac{dV_0}{dV_r} \,.
\end{equation}
Computing the transformed volume element $dV_r$ requires solving the geodesic equations associated with the spacetime metric.

The ratio $dV_0/dV_r$ quantifies the gravitational focusing of the incident beam, capturing how spacetime curvature compresses or expands the initial volume element along the trajectory. The metric of a non-rotating\footnote{Or slowly rotating.} compact star is given by
\begin{equation}
ds^2 = -g_{tt}(r)\,dt^2 + g_{rr}(r)\,dr^2 + r^2\,d\Omega^2 \,,
\end{equation}
where $g_{tt}$ and $g_{rr}$ are functions of the radial coordinate $r$.

The geodesic equations for a particle starting far from the star and propagating with impact parameter $b$ can be written by exploiting the spherical symmetry of the system to restrict the motion to the equatorial plane, which we identify with $\theta = \pi/2$:
\begin{equation}
\begin{split}
\frac{dt}{d\lambda} & = \frac{\mathcal{E}}{g_{tt}}\,,\qquad 
\frac{dr}{d\lambda} = \pm \sqrt{
\frac{\mathcal{E}^2}{g_{tt} g_{rr}}
- \frac{1}{g_{rr}} \left( 1 + \frac{\mathcal{L}^2}{r^2} \right)
}\,,\\ 
\frac{d\phi}{d\lambda} & = \frac{\mathcal{L}}{r^2}\,,\qquad
\frac{d\theta}{d\lambda}  = 0\,,
\end{split}
\label{eq:geodesics}
\end{equation}
where $\mathcal{E} \equiv E_\chi / m_\chi$ is the conserved energy per unit mass far from the star and $\mathcal{L}$ is the conserved angular momentum per unit mass. The impact parameter $b$ labels each trajectory at infinity and is related to $\mathcal{L}$ through $\mathcal{L} \equiv -b\,p_\chi/m_\chi$.
The energy is defined in the asymptotic frame and does not correspond to the local energy at each point along the trajectory. The sign in the radial equation is negative while the particle approaches the star and becomes positive after reaching its minimum radius, determined by the condition
\[
\frac{\mathcal{E}^2}{g_{tt}} = 1 + \frac{\mathcal{L}^2}{r^2}\,.
\]
This turning point corresponds to the vanishing of the radial component of the momentum, marking the balance between gravitational attraction and angular momentum.

The initial conditions are chosen as
\begin{equation}
\begin{aligned}
t_0 &= 0\,, \qquad
r_0 = \sqrt{L^2 + b^2}\,,\\
\phi_0 &= \pi - \sin^{-1}\!\left( \frac{b}{L} \right)\,, \qquad
\theta_0 = \frac{\pi}{2}\,.
\end{aligned}
\end{equation}
The volume element associated with the initial beam is evolved along the geodesics by parallel transporting
$dV_0 \equiv *_{3D}(dt \wedge dA_\perp)$,
where $*_{3D}$ denotes the pull-back onto the three-dimensional spatial hypersurface. To describe this evolution, we select a congruence of geodesics and use an auxiliary parameter $\tilde{b}$ as a label that remains constant along each trajectory and is equal to the impact parameter $b$ for each geodesic. The parameters $(\tilde{b},t)$ therefore span a two-dimensional surface transported by the geodesic flow, corresponding to the directions along which gravitational focusing distorts the beam.

To construct a three-dimensional volume element, an additional transverse direction is required. This direction is provided by an auxiliary angular coordinate, denoted $\theta_{yz}$, which parametrizes rotations of the congruence around the beam axis. Since both the incoming flux and the spacetime metric are symmetric under such rotations, $\theta_{yz}$ does not enter the geodesic equations and simply labels equivalent copies of the equatorial congruence.

With this interpretation, the transported volume can be visualized as a set of rings with radius
$y \equiv r \sin\phi$
(which corresponds to the impact parameter $b$ at infinity), whose area is spanned by the directions $(\tilde{b},t)$. An infinitesimal segment of such a ring is obtained by noting that $\theta_{yz}$ is orthogonal to the other variables, leading to the volume element
\begin{equation}
    dV_r = *_{3D}(\sigma_{\tilde{b}} \wedge \sigma_t \wedge \sigma_{\theta_{yz}})
      = *_{3D}(r \sin\phi \, \sigma_{\tilde{b}} \wedge \sigma_t \wedge d\theta_{yz})\,.
\end{equation}
Geometrically, this construction tracks how an initially flat area associated with the beam is stretched or compressed as the congruence of geodesics bends through the stellar spacetime, thereby determining the local density of DM particles.

The area $\sigma_{\tilde{b}} \wedge \sigma_t$ is obtained by defining the transformation from the parameters $(\tilde{b}, t)$ to the spherical coordinates $(r, \phi)$. Points along each geodesic are described by
$X^\mu \equiv (t, r(\lambda, \tilde{b}), \theta, \phi(t, \tilde{b}))$.
Since the motion is restricted to the equatorial plane, only $r$ and $\phi$ depend on $t$ and $\tilde{b}$. The corresponding four-velocity is given by
\begin{equation}
u^\mu(t, \tilde{b}) = \frac{\partial X^\mu}{\partial t}.
\end{equation}
To compute the area, we introduce the vectors
\begin{equation}
\hat{e}_{\tilde{b}}^\mu = \frac{dX^\mu}{d\tilde{b}},
\qquad
\hat{e}_{t}^\mu = \frac{dX^\mu}{dt},
\end{equation}
which span the infinitesimal patch transported along the geodesic congruence.
A two-dimensional metric $\tilde{g}^{2D}_{ij}$ is then defined by taking the inner products of these vectors, projected onto the corresponding two-dimensional subspace through the reduced metric
\begin{equation}
\hat{g}_{\mu\nu} \equiv g_{\mu\nu} + u^t_\mu u^t_\nu + u^\theta_\mu u^\theta_\nu
    = \mathrm{diag}(0, g_{rr}, 0, r^2),
\end{equation}
which is appropriate for motion restricted to the equatorial plane ($\theta = \pi/2$).
This projection tracks how the patch spanned by $(\tilde{b},t)$ is stretched or compressed by the geodesic flow, yielding the physical area. The resulting two-dimensional metric is therefore
\begin{equation}
\tilde{g}^{2D}_{ij} = \hat{g}_{\mu \nu} \hat{e}^\mu_i \hat{e}^\nu_j \,.
\end{equation}
The area associated with the deformation along the geodesic congruence is then given by
\begin{equation}
\label{eq:area}
\begin{split}
\sigma_{\tilde{b}} \wedge \sigma_t
= \sqrt{|\tilde{g}^{2D}|} \, d\tilde{b} \wedge dt
= r \, \sqrt{g_{rr}} \left| r_t \phi_{\tilde{b}} - r_{\tilde{b}} \phi_{t} \right| \, d\tilde{b} \wedge dt\,,
\end{split}
\end{equation}
where we have introduced the shorthand notation $X_\alpha \equiv \partial X / \partial \alpha$.

Combining the expressions for $dN$, the transported volume element $dV_r$, and the local interaction rate $\Omega^-_{v_e}$, we obtain the total beam-induced rate inside the star. This yields
\begin{equation}
\label{eq:capture_rate_gen_LAB}
\begin{split}
C &= \int dn_\chi(T^\infty_\chi, r, \phi) \,
      \Omega^-_{v_e}\big(\omega(T^\infty_\chi, r)\big) \, dV
\\
&= \int dV \int dT^\infty_\chi
     \frac{d\Phi}{dT^\infty_\chi} \,
     \frac{dV_0}{dV_r} \,
     \Omega^-_{v_e}\big(\omega(T^\infty_\chi, r)\big) \, ,
\end{split}
\end{equation}
where
\begin{equation}
\frac{dV_0}{dV_r}
= \frac{b \, r_0 \, \left| r_t \phi_{\tilde{b}} - r_{\tilde{b}} \phi_{t} \right|_0}
       {r^2 \, \sin\phi \, \sqrt{g_{rr}} \,
        \left| r_t \phi_{\tilde{b}} - r_{\tilde{b}} \phi_{t} \right|} \, ,
\end{equation}
is the ratio between the volume element evaluated far from the star and that at position $\vec{r}$ inside the star. Following~\cite{HoefkenZink:2024hor}, the interaction rate can be written as
\begin{align}
\label{eq:Omega_gen}
\Omega^-_{v_e}(\omega, r)
    &= \sqrt{g_{tt}(r)}
       \int_{0}^{v_e} dv \, \frac{d\sigma}{dv} \, \omega \, n_T(r)
       = \sqrt{g_{tt}(r)} \, \omega \, n_T(r) \, \sigma_\mathrm{cap}\, ,
\end{align}
where $n_T(r)$ denotes the number density of target particles at radius $r$, and $\sigma_\mathrm{cap}$ is the cross section restricted to the capture condition, namely that the outgoing particle velocity lies below the local escape speed $v_e$. This expression is not valid when the target particles cannot be treated as being at rest, but instead follow a Fermi--Dirac distribution. A more general treatment appropriate for degenerate targets is presented in \Cref{sec:degenerated}.

Since we are interested in heating rather than gravitational binding, we do not distinguish which particles become bound to the star. Accordingly, in the heating calculation we replace $\sigma_\mathrm{cap}$ by the total cross section in the relevant regime, and substitute $\Omega^-_{v_e}(\omega, r)$ with $\Omega^- \equiv \Omega^-_{v_e = c}$, so that we do not track whether the particle is captured or not and consider any outgoing velocity up to $c$. The factor $\sqrt{g_{tt}}$ accounts for the conversion between proper time and coordinate time when the rate is measured from infinity.

\begin{figure}
    \centering
    \includegraphics[width=0.48\linewidth]{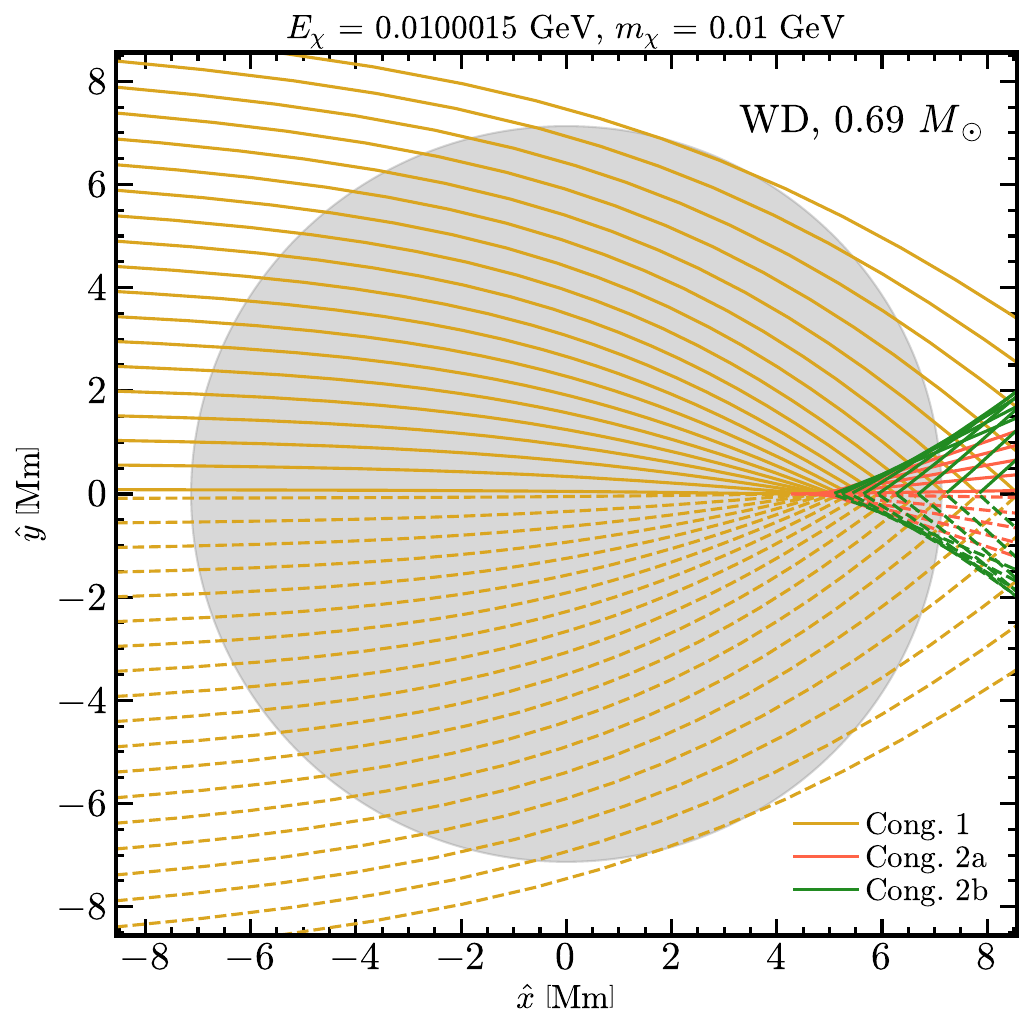}
    \includegraphics[width=0.48\linewidth]{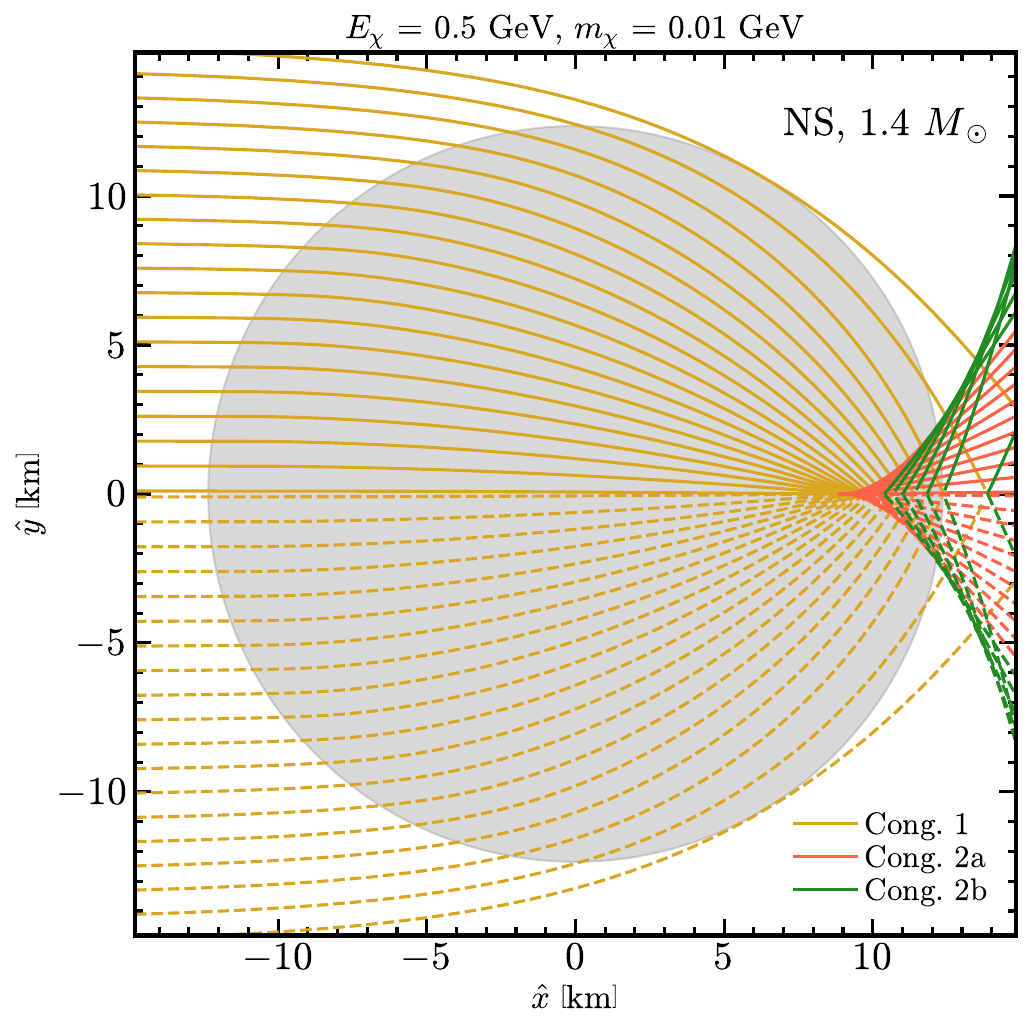}
    \caption{
    Example of geodesic trajectories used to determine the number of particles per unit volume for a WD (left) and an NS (right).
    Each curve represents the trajectory of a BBDM particle with a different impact parameter $b$, originating in the upper
    half--plane and propagating toward the compact object (grey disk).
    Region~1 (gold) corresponds to single, non-degenerate trajectories.
    Regions~2a (orange) and~2b (green) represent overlapping congruences of geodesics that both traverse the lower
    half--plane.
    The overlap of these families defines the multi--stream region relevant for computing $dN$.
    }
    \label{fig:amp_regions}
\end{figure}

A subtlety must be taken into account at this stage. By construction, geodesic congruences are defined such that each spatial point is traversed by a single geodesic. However, at sufficiently low energies there exist regions of the star where multiple families of geodesics intersect the same spatial domain. In this case, the quantity $dV_0/dV_r$ must be computed separately for each contributing family. In principle, many such families and regions may arise, each contributing independently to the local particle density. In practice, for energies that are not extremely low, it is sufficient to consider three  families of congruences, as illustrated in \Cref{fig:amp_regions}, where they are labeled as ``Cong.~1'', ``Cong.~2a'', and ``Cong.~2b''.

Restricting to the equatorial plane and considering geodesics originating in the upper half--plane, family~1 corresponds to trajectories that remain entirely in the upper half--plane inside the star. Families~2a and~2b correspond to trajectories that cross into the lower half--plane. This region is degenerate in the sense that there exists a minimum impact parameter $b_{\mathrm{min}}$ such that geodesics with $0 < b < b_{\mathrm{min}}$ and those with $b_{\mathrm{min}} \le b < b'$ (for some value $b'$) each cover the entire region of the lower half--plane accessible to trajectories originating in the upper half--plane. Geodesics with $b > b'$ do not reach the lower half--plane.
Each family of congruences will be treated separately.

Exploiting the symmetry of the problem under reflection across the equatorial plane, we restrict our analysis to the upper half--plane. Contributions from trajectories entering the lower half--plane can be projected onto it, ensuring that all relevant regions are properly accounted for. As a result, the total contribution takes the form
\begin{equation}
\left[ \frac{dV_0}{dV_r} \right]
\;\to\;
\left[ \frac{dV_0}{dV_r} \right]_1
+ \left[ \frac{dV_0}{dV_r} \right]_{2a}
+ \left[ \frac{dV_0}{dV_r} \right]_{2b} \, ,
\end{equation}
where the contributions from families~2a and~2b are set to zero whenever their corresponding geodesics do not cover the volume under consideration. Whenever there is just one geodesic passing through a point in space, there is just one contribution taken into account.

The resulting sum accounts for the full multi--stream structure of the flow, which enhances the local particle density and must be included consistently in the computation of $dN$ and, consequently, in all rate--related quantities.
Once the full geodesic structure has been incorporated into the local number density, the same geometric information
can be used to compute not only capture, but also the energy deposited by trajectories that scatter inside the star,
independently of whether they become gravitationally bound.

Beyond gravitational binding, BBDM can deposit energy even when the particles do not become bound. Operationally, the total power is
obtained with the same kernel structure as $C$, multiplied by the energy deposited per collision, with additional
factors accounting for the frame in which the heating is evaluated. Since energies cannot be summed across different
local frames, all contributions are transported to the surface frame of the star. Therefore, the full expression for
energy deposition is
\begin{equation}
\begin{split}
\dot E_\chi &= 2\pi \int d\rho \, dz \, dT^\infty_\chi \, d\Pi \, \rho \,
\sqrt{\frac{g_{rr} (\rho, z)}{g_{tt}^s}} \,
X(\rho, z)
\frac{d\Phi}{dT^\infty_\chi}
\frac{d\Omega^-}{d\Pi}
E_k (\Pi, v_e) \,,\\
&= 2\pi \int d\rho \, dz \, dT^\infty_\chi \, d\Pi \, \rho \,
\sqrt{\frac{g_{rr} (\rho, z) \, g_{tt} (\rho, z)}{g_{tt}^s}} \,
n_T (\rho, z) \, \omega
\left(1 + \frac{T^\infty_\chi}{m_\chi}\right)
X_V(\rho, z)
\frac{d\Phi}{dT^\infty_\chi}
\frac{d\sigma}{d\Pi}
E_k (\Pi, v_e) \,,
\end{split}
\end{equation}
where the integration is performed in cylindrical coordinates $(\rho, z, \phi)$, with $\phi$ integrated out by symmetry.
The variable $\Pi$ denotes the parameter space relevant for the cross section, depending on whether we are in the DIS,
RES, or elastic regime. The escape velocity is $v_e$, and the superscript $s$ refers to quantities
evaluated at the stellar surface. The factor $\sqrt{g_{tt}(\rho, z)/g_{tt}^s}$ accounts for the conversion between proper
time and the surface frame. The geometric factor
\begin{equation}
X_V(\rho, z) \equiv \frac{dV_0}{dV_r}\,,
\end{equation}
encodes the deformation of the incoming flux due to the geodesic mapping, and
\begin{equation}
E_k (\Pi, v_e) =
\sqrt{\frac{g_{tt} (\rho, z)}{g_{tt}^s}}
\begin{cases}
E_\chi (r) - m_\chi &\text{if } v \le v_e\,, \\
q_0 (\Pi) &\text{if } v > v_e\,,
\end{cases}
\end{equation}
where $E_\chi$ is the local energy of the particle at radius $r=\sqrt{\rho^2+z^2}$, and $q_0(\Pi)$ is the transferred
energy at the scattering point. If the particle escapes, only the transferred energy contributes; if it becomes gravitationally bound, the
entire kinetic energy is eventually deposited. The factor
$\sqrt{g_{tt} (\rho, z)/g_{tt}^s}$ ensures that energies are consistently measured in the frame of the stellar surface.

This final expression is valid when we consider target particles at rest. If they have a Fermi-Dirac distribution, which is true for any target particle in a NS or for electrons in a WD, we need to compute the interaction rate in a more involved way, as shown in the next subsection.

\subsection{Interaction rate of degenerate matter}
\label{sec:degenerated}

We now recompute \Cref{eq:Omega_gen} using an appropriate target distribution, considering the process
$\chi (k_1) + p (p_1) \to \chi (k_2) + X (p_2)$,
where the final-state particle $X$ depends on the interaction regime under consideration: $p$ for elastic scattering, and a generic hadronic state $X$ for deep inelastic scattering (DIS). Since the target medium is degenerate, the number density is replaced by
\begin{equation}
\begin{split}
n_T &\;\to\; d^3 p_{p_1} \, \frac{g_s}{(2\pi)^3} \, n_F\!\left(E_{p_1}\right) \,,\\
&=\; \frac{p_{p_1} E_{p_1} \, dE_{p_1} \, d\cos\theta_{\text{uw}}}{2\pi^2}
\, n_F\!\left(E_{p_1}\right)\,,
\end{split}
\end{equation}
where $\theta_{\text{uw}}$ is the angle between the trajectories of the incoming DM particle $\chi$ and the target particle, and $n_F(E)$ denotes the Fermi--Dirac distribution evaluated at the stellar temperature.

It is convenient to transform the integration variable from $\cos\theta_{\text{uw}}$ to the Mandelstam variable $s$. The corresponding Jacobian is
\begin{equation}
\begin{split}
\left|\frac{d\cos\theta_{\text{uw}}}{ds}\right|
= \frac{1}{2 p_{p_1}}
\sqrt{\frac{g_{tt}(r)}
{\left(E^\infty_\chi\right)^2 - g_{tt}(r)\, m_\chi^2}}\,,
\end{split}
\end{equation}
where $E^\infty_\chi$ is the total energy of the DM particle measured far from the star. This Jacobian allows us to rewrite the integration over $\cos\theta_{\text{uw}}$ as an integration over $s$.

In addition, the relative velocity factor $\omega$ appearing in \Cref{eq:Omega_gen} must be replaced by $\left|\vec{w} - \vec{u}_T\right|$, where $\vec{w}$ and $\vec{u}_T$ are the velocities of the DM particle and the target, respectively, evaluated in the center-of-momentum frame. The differential cross section is also computed in this frame. We find
\begin{equation}
\begin{split}
\left|\vec{w} - \vec{u}_T\right|
&= \frac{2s\,\sqrt{s^2 - 2\left(m_\chi^2 + m_T^2\right)s
+ \left(m_\chi^2 - m_T^2\right)^2}}
{s^2 - \left(m_T^2 - m_\chi^2\right)^2}\,,\\
&\equiv \frac{2s\, g_s(m_\chi)}{\beta_s(m_\chi)}\,,
\end{split}
\end{equation}
where $m_T$ is the target mass, and the functions $g_s$ and $\beta_s$ are introduced as shorthand notation.
Following~\cite{Bell:2020jou}, we include the factor $\zeta \equiv n_T / n_\mathrm{free}$, with
$n_\mathrm{free} = \left[\mu_T \left(2 m_T + \mu_T\right)\right]^{3/2} / (3\pi^2)$,
where $\mu_T$ is the chemical potential of the target. This factor accounts for the fact that the number density obtained from the equation of state does not exactly coincide with that derived from the Fermi--Dirac distribution used above.

To perform the integration of the differential cross section, we must distinguish between the different interaction regimes. In this work, we consider the same set of processes that will be discussed in \Cref{sec:examples}: elastic scattering between DM and nucleons, and deep inelastic scattering (DIS).

\subsubsection{Elastic interactions}

In the elastic regime, the computation can be performed in the center-of-momentum frame, where the differential cross section takes the form
\begin{equation}
\begin{split}
\frac{d \sigma}{d \cos \theta_\mathrm{cm}} = \frac{\langle \mathcal{M}^2 \rangle}{32 \pi \, s}\,.
\end{split}
\end{equation}
It is convenient to change variables to the Mandelstam variable $t$, which is better suited for these computations. The relation between $t$ and $\cos \theta_\mathrm{cm}$ is given by
\begin{equation}
\begin{split}
\frac{d \cos \theta_\mathrm{cm}}{dt}
&= \frac{2s}{s^2 - 2s\left(m_T^2 + m_\chi^2\right)
+ \left(m_T^2 - m_\chi^2\right)^2} \\
&\equiv \frac{2s}{g_s^2(m_\chi)}\,.
\end{split}
\end{equation}
In addition, Pauli blocking must be taken into account by including the factors
$n_F\!\left(E_{p_1}\right)\left[1 - n_F\!\left(E_{p_2}\right)\right]$,
which account for the occupation of the initial target state and the availability of the final target state in degenerate matter.

Combining all these ingredients, the interaction rate in the elastic regime can be written as
\begin{equation}
\begin{split}
\Omega^-_\mathrm{elas}(r)
=&\;\frac{3}{64 \pi^2}
   \frac{n_T}{\left[\mu_T \left(2 m_T + \mu_T\right)\right]^{3/2}}
   \frac{g_{tt}(r)}{\sqrt{\left(E^\infty_\chi\right)^2 - g_{tt}(r)\, m_\chi^2}} \\
&\times \int dE_{p_1} \, ds \, dt \, d\phi_{k_1 p_1}
   \frac{\langle \mathcal{M}^2 \rangle}{g_s(m_\chi)}
   \frac{s \, E_{p_1}}{\beta_s(m_\chi)} \\
&\times n_F\!\left(E_{p_1}\right)
   \left[1 - n_F\!\left(E_{p_2}\right)\right]\,,
\end{split}
\end{equation}
where $\phi_{k_1 p_1}$ denotes the polar angle between the incoming DM particle and the target momenta in spherical coordinates. This dependence is required because the energy of the outgoing target particle depends on this angle, and it must therefore be retained in order to properly account for Pauli blocking. We note that this angular dependence has often been neglected in previous studies by assuming that all particle momenta lie on the same plane. However, this assumption does not hold in general, and we explicitly include this dependence in our calculation.

The kinematic limits of the Mandelstam variable $s$ are given by
\begin{equation}
\begin{split}
s_{\mathrm{max}/\mathrm{min}}
= \pm\, 2 \sqrt{\left(E_{k_1}^2 - m_\chi^2\right)
                 \left(E_{p_1}^2 - m_T^2\right)}
  + 2 E_{k_1} E_{p_1}
  + m_T^2 + m_\chi^2\,.
\end{split}
\end{equation}
For the Mandelstam variable $t$ the integration limits are
\begin{equation}
\begin{split}
t_{\mathrm{min}} &= -\frac{g_s^2\!\left(m_\chi\right)}{s}\,,\qquad
t_{\mathrm{max}} = 0\,.
\end{split}
\end{equation}
Finally, the target energy $E_{p_1}$ is integrated from the target mass $m_T$ up to the chemical potential $\mu_T$, while the angular variable $\phi_{k_1 p_1}$ ranges from $0$ to $2\pi$.

\subsubsection{DIS regime}

In the DIS regime, the cross section can be expressed in terms of the standard DIS variables,
\begin{equation}
dv \, \frac{d\sigma}{dv} \;\to\; dx \, dy \, \frac{d^2 \sigma}{dx \, dy}\,,
\end{equation}
where $x \equiv Q^2 / (2 p_1 \cdot q)$ is the Bjorken scaling variable and
$y \equiv p_1 \cdot q / (p_1 \cdot k_1)$ is the inelasticity.

Since the outgoing hadronic state corresponds to a jet rather than a single hadron, no specific final-state hadron is identified. Moreover, the relevant energies in the DIS regime are sufficiently large that Pauli blocking effects are not expected to be important. Accordingly, we do not include Pauli blocking for the outgoing hadronic states.

The resulting expression for the interaction rate in the DIS regime is
\begin{equation}
\begin{split}
\Omega^-_\mathrm{DIS}(r)
=&\;\frac{3\, n_T}{2\left[\mu_T \left(2 m_T + \mu_T\right)\right]^{3/2}}
   \frac{g_{tt}(r)}{\sqrt{\left(E^\infty_\chi\right)^2 - g_{tt}(r)\, m_\chi^2}} \\
&\times \int dE_{p_1} \, ds \, dx \, dy \;
   E_{p_1}\,
   \frac{d^2 \sigma}{dx \, dy}
   \frac{s \, g_s(m_\chi)}{\beta_s(m_\chi)}
   \, n_F\!\left(E_{p_1}\right)\,,
\end{split}
\end{equation}
where the integration limits for $E_{p_1}$ and $s$ are the same as in the elastic case. The integration ranges for the DIS variables $x$ and $y$ follow the standard DIS kinematics. In this regime, the dependence on the angle $\phi_{k_1 p_1}$ is no longer required, since the energy of the outgoing hadronic system is not explicitly reconstructed.

\subsection{Interaction roof and geometric limit}

The geometric limit corresponds to the regime in which all particles traversing the star are captured. In the context of boosted DM, we also define an \emph{interaction roof} (IR) as the regime in which all particles interact with the star and therefore deposit energy into it, regardless of whether they become gravitationally bound. To compute the total energy deposition rate in the IR, we must combine two ingredients:  
(1) the total number of particles passing through the star per unit time, and  
(2) the average energy deposited per interaction.  
Both quantities are evaluated separately for each energy component of the incoming flux.

In what follows, we adopt a simplifying assumption: the interaction occurs at the stellar surface. This approximation effectively neglects the redshift between the interaction point and the surface, since we take the interaction to occur directly at that location.

To determine the number of particles crossing the compact object per unit time, we compute the maximum impact parameter $b_\mathrm{max}$ for which a particle with fixed energy at infinity reaches the stellar surface. This quantity is obtained by setting the radial geodesic equation $dr/d\lambda$ to zero at $r = R_*$ and assuming a Schwarzschild metric, since the motion occurs entirely outside the star. The resulting expression is
\begin{equation}
b_\mathrm{max}
= \sqrt{
\frac{\left[(T^\infty_\chi)^2 + 2 m_\chi T^\infty_\chi\right] R_* + m_\chi^2 r_s}
     {\left[(T^\infty_\chi)^2 + 2 m_\chi T^\infty_\chi\right]\left(R_* - r_s\right)}
     }\, R_* \, .
\end{equation}
The total number of particles traversing the star is then obtained by considering the flux through a cross-sectional area $\pi b_\mathrm{max}^2$. In addition, we include a factor of $1/\sqrt{g_{tt}(R_*)}$ to convert the rate to the proper time measured at the stellar surface. This gives
\begin{equation}
\frac{dN_\chi}{dt}
= \int dT^\infty_\chi \,
  \frac{d\Phi}{dT^\infty_\chi} \,
  \left(\pi b_\mathrm{max}^2\right)
  \frac{1}{\sqrt{g_{tt}(R_*)}} \, .
\end{equation}

Next, we include the mean energy transferred per interaction, given by
\begin{equation}
\langle q_0 \rangle
= \frac{m_\chi + T^\infty_\chi}{\sqrt{g_{tt}(R_*)}}
  \frac{\int d\Pi \, y \, \frac{d \sigma}{d\Pi}}
       {\sigma_\mathrm{tot}(T^\infty_\chi)} \, ,
\end{equation}
where $q_0 \equiv (m_\chi + T^\infty_\chi)\, y$, with $y$ equal to the inelasticity, $\Pi$ represents the variables that describe the parameter space of the interaction, and the factor
$1/\sqrt{g_{tt}(R_*)}$ accounts for the redshift of the deposited energy to the stellar surface. In case we are dealing with degenerate matter targets, the previous equation takes a different form, since we need to perform the integration in the distribution of targets and divide by the number density of them, $n_T$:
\begin{equation}
\langle q_0 \rangle
= \frac{m_\chi + T^\infty_\chi}{4 \, \pi^2 \, \sqrt{g_{tt}(R_*)} \, n_T} \times \sqrt{\frac{g_{tt}(r)}
{\left(E^\infty_\chi\right)^2 - g_{tt}(r)\, m_\chi^2}} \times
  \frac{\int E_{p_1} dE_{p_1} \, ds \, d\Pi \, y \, \frac{d \sigma}{d\Pi} \times n_F\!\left(E_{p_1}\right) \times \left(1 - n_F\!\left(E_{p_2}\right)\right)}
       {\sigma_\mathrm{tot}(T^\infty_\chi)} \, ,
\end{equation}
where the procedure and the limits of integration are the same as stated in the previous subsection.

The total energy deposition rate in the interaction-roof regime is therefore
\begin{equation}
\langle \dot{E}_\chi \rangle_\mathrm{IR}
= \int dT^\infty_\chi \,
  \frac{d\Phi}{dT^\infty_\chi} \,
  \left(\pi b_\mathrm{max}^2\right)
  \frac{\langle q_0 \rangle}{\sqrt{g_{tt}(R_*)}} \, .
\end{equation}

If multiple scatterings occur such that the entire kinetic energy of the particle is deposited inside the star, the system enters the true geometric-limit regime. In this case, $\langle q_0 \rangle$ must be replaced by the total available kinetic energy,
$(m_\chi + T^\infty_\chi)/\sqrt{g_{tt}(R_*)} - m_\chi$, yielding
\begin{equation}
\langle \dot{E}_\chi \rangle_\mathrm{GL}
= \int dT^\infty_\chi \,
  \frac{d\Phi}{dT^\infty_\chi} \,
  \left(\pi b_\mathrm{max}^2\right)
  \frac{\left(1 - \sqrt{g_{tt}(R_*)}\right)m_\chi + T^\infty_\chi}
       {g_{tt}(R_*)} \, .
\end{equation}

\subsection{Optical factor}

As particles propagate through the star, sufficiently strong interactions can reduce the number of particles reaching a given point inside the stellar interior. This effect can be described in terms of absorption and is accounted for by introducing an optical factor. In this subsection, we show how to compute this factor under the assumption of single-scattering heating, i.e., we assume that once a particle interacts with the stellar material, it cannot undergo further interactions.

The evolution of the number of particles along a fixed geodesic inside the star is governed by
\begin{equation}
\frac{dN_\chi}{dt^\mathrm{loc}} = -\frac{\Omega^-}{\sqrt{g_{tt}(r)}}\, N_\chi \,,
\end{equation}
where $N_\chi$ denotes the number of particles crossing a given point and $t^\mathrm{loc}$ is the time measured locally by an observer at radius $r$. The factor $1/\sqrt{g_{tt}(r)}$ ensures that $\Omega^-$ is evaluated with respect to the local proper time rather than the time measured at infinity. The solution to this equation is
\begin{equation}
\frac{N_\chi}{N_\chi^{(0)}}
= \exp\!\left(- \int dt^\mathrm{loc}\,
\frac{\Omega^-}{\sqrt{g_{tt}(r)}} \right) \,,
\end{equation}
this expression is precisely the optical factor  $\eta$.  Rewriting the integral in terms of the global time coordinate,
\begin{equation}
\eta
= \exp\!\left(- \int dt\, \Omega^- \right)\,.
\end{equation}
The integral can be conveniently expressed in terms of the polar angle $\phi$ along the equatorial plane. Using the geodesic equations given in \Cref{eq:geodesics}, we have
\begin{equation}
\begin{split}
\frac{dt}{d\phi}
&= \frac{(m_\chi + T^\infty_\chi)\, r^2}{g_{tt}(r)\, J(b,T_\chi^\infty,m_\chi)}
\end{split}
\end{equation}
with $J(b,T_\chi^\infty, m_\chi) = b\,\sqrt{\left(T^\infty_\chi\right)^2 + 2 m_\chi T^\infty_\chi}$.

The choice of $\phi$ instead of $r$ as the integration variable is motivated by the fact that the geodesic inside the star is single-valued in $\phi$, while this is not generally the case for $r$. The exponent appearing in the optical factor then becomes
\begin{equation}
\int dt\, \Omega^-
= \int_{\mathcal{C}} d\phi \, \frac{dt}{d\phi}\, \Omega^- \,,
\end{equation}
where $\mathcal{C}$ denotes the trajectory along a given geodesic. This expression can be written explicitly as
\begin{equation}
\begin{split}
\int dt\, \Omega^-
= \frac{m_\chi + T^\infty_\chi}
        {J(b,T_\chi^\infty,m_\chi)}
   \int_{\phi^*}^{\phi(r)} d\phi'\,
   \left[r'(\phi')\right]^2
   \int d\Pi \, \frac{d\sigma}{d\Pi}\,
   \omega \, n_T\!\left(r'(\phi')\right)\,,
\end{split}
\end{equation}
here, $\phi(r)$ is the polar angle corresponding to the position at which the optical factor is evaluated, and $\phi^*$ is the angle at which the geodesic enters the star. The function $r'(\phi')$ gives the radial position along the geodesic, and $\Pi$ denotes the phase space of the cross section, which depends on the interaction channel under consideration. Contributions from all relevant interaction processes must be included. We emphasize that the integration is not restricted by the escape velocity, since we are interested in any interaction, not only those leading to capture. Furthermore, the previous expression can be transformed and integrated in the proper time of the particle going through the star.

For geodesics that cross the $\hat{x}$ axis, the optical factor must be included in the volume amplification, as multiple geodesics contribute to the same spatial point inside the star. Absorption effects must therefore be incorporated consistently into the total amplification. The resulting expression is
\begin{equation}
X_V^\mathrm{opaque}(\rho, z)
\equiv \left[ \frac{dV_0}{dV_r} \right] \times \eta(\rho, z)
\;\to\;
\left[ \frac{dV_0}{dV_r} \right]_1 \eta(\rho, z)_1
+ \left[ \frac{dV_0}{dV_r} \right]_{2a} \eta(\rho, z)_{2a}
+ \left[ \frac{dV_0}{dV_r} \right]_{2b} \eta(\rho, z)_{2b}\,,
\end{equation}
where the subscripts label the different families of geodesics contributing to the same spatial region.

\subsection{Isotropic fluxes}

The heating formalism derived for a directed flux can be extended to more general flux configurations. Fluxes that are isotropic far from the star, such as CRBDM~\cite{Bringmann:2018cvk} or even cosmic rays, can also be described within our formalism, provided that the compact object is spherically symmetric. To demonstrate this, we compute the contribution to the number density of particles associated with a fixed impact parameter $b$.

For a jet with a fixed direction at infinity, the perpendicular area traversed by particles with impact parameter between $b$ and $b + db$ is
\begin{equation}
dA^\mathrm{jet}_\perp = 2 \, \pi \, b \, db\,.
\end{equation}
We now show that the same area is obtained for an isotropic flux. If this is the case, and the compact star is spherically symmetric, then all trajectories with a given impact parameter can be rotated to originate from a single direction without affecting the capture or heating rates. The computation then becomes mathematically equivalent to that of a directed jet.

Let us consider a sphere of radius $R$, with $R \gg R_*$. We evaluate the fraction of an isotropic flux passing through the sphere that corresponds to a given impact parameter. The direction of the flux at any point on the sphere can be parametrized by introducing a local coordinate system centered at that point, with the $\hat{z}$ axis pointing toward the star. Since the sphere is located far from the star, only trajectories with small polar angles $\theta \simeq 0$ in this frame contribute. The impact parameter can then be written as
\begin{equation}
\begin{split}
b &= R \, \sin \theta\,,\\
db &= R \, \cos \theta \, d\theta \simeq R \, d\theta\,.
\end{split}
\end{equation}

The large separation between the sphere and the star further justifies treating the incoming flux lines as perpendicular to the surface of the sphere. The fraction of the isotropic flux corresponding to trajectories with polar angle between $\theta$ and $\theta + d\theta$, i.e. with fixed impact parameter $b$, is then given by
\begin{equation}
\mathcal{F}_b
= \frac{\sin \theta \, d\theta}{\int_0^\pi \sin \theta \, d\theta}\,.
\end{equation}
Using the relation between the polar angle $\theta$ and the impact parameter $b$, this fraction can be expressed as
\begin{equation}
\mathcal{F}_b
= \frac{b \, db}{2 \, R^2 \cos \theta}
\simeq \frac{b \, db}{2 \, R^2}\,,
\end{equation}
where in the last step we have used $\cos \theta \simeq 1$, valid for trajectories originating far from the star.

The total perpendicular area traversed by the component of the isotropic flux with impact parameter $b$ is obtained by multiplying this fraction by the total area of the sphere,
\begin{equation}
dA^\mathrm{iso}_\perp
= 4 \, \pi \, R^2 \, \mathcal{F}_b \,.
\end{equation}
This yields
\begin{equation}
dA^\mathrm{iso}_\perp
= 2 \, \pi \, b \, db,
\end{equation}
which coincides exactly with the perpendicular area associated with a directed jet $dA^\mathrm{jet}_\perp$. Therefore, an isotropic flux can be treated as an effectively single-directed flux, provided that the star is spherically symmetric, so that trajectories with the same impact parameter can be rotated without altering their contribution to the capture rate or to the heating of the star.

\section{Examples}
\label{sec:examples}

In this section, we present a representative example of the heating of compact stars by DM particle beams.
We begin by specifying the particle physics framework and interaction channels relevant for energy deposition inside
compact objects. We then construct the corresponding boosted DM flux from astrophysical sources and apply
the formalism developed in the previous sections to compute the resulting heating rates for white dwarfs and neutron
stars. The results illustrate the role of different interaction regimes and optical depths, including the optically
thin limit, the interaction roof, and the geometric limit, for a set of benchmark models.

All multidimensional integrals are evaluated using the \texttt{vegas} integrator implemented in C++~\cite{Lepage:1977sw,Lepage:2020tgj}.
For the parton distribution functions, we employ the \textit{CT18NNLO} PDF set~\cite{Buckley:2014ana}.
The geodesic structure and associated amplification factors are computed using the \texttt{PyGRO} package~\cite{pygro2025},
which we use to generate maps as functions of position, coordinate time $t$, and impact parameter $b$.
White dwarf profiles are generated using our own numerical implementation, while neutron star profiles are obtained from a
modified version of the \texttt{CompactObject} code~\cite{Huang:2023grj,Huang:2024ewv,Huang:2024rfg,Huang:2024rvj},
adapted to output the radius, mass, pressure, and energy density at each step of the TOV integration.

\subsection{Interactions}

The computation of the capture and heating rates inside stars requires specifying the microscopic DM--SM interaction.
We consider fermionic DM interacting through an axion-like-particle (ALP) mediator (a pseudoscalar) and restrict ourselves to interactions between DM and
quarks,
\begin{equation}
\label{eq:Lag}
\mathcal{L}_{a} = \sum_{q} g_{q a} \partial_\mu a \overline{q} \gamma^\mu \gamma^5 q
+ g_D \partial_\mu a \overline{\chi} \gamma^\mu \gamma^5 \chi\,,
\end{equation}
where the couplings $g_{q a}$ and $g_{D}$ have both units of inverse energy. These kinds of models are suitable for high energy interaction searches, since their cross sections scale with powers of the transferred momentum, so that the interactions are suppressed at low energies. This is precisely a model that should be studied with boosted DM, since traditional direct detection experiments have a very low sensitivity to them.

For elastic interactions of the form $\chi(k_1) + N(p_1) \to \chi(k_2) + N(p_2)$, the amplitude mediated
by an ALP is written as~\cite{DeMarchi:2025uoo}
\begin{equation}
\begin{split}
i \mathcal{M} = i g_D q_\mu q_\nu  \overline{u} (k_2) \gamma^\mu \gamma^5 u (k_1) \frac{i}{Q^2 + m_a^2} \frac{2 \, m_N}{m_u + m_d} \sum_q g_{q a} G_{5, N}^q (Q^2) \overline{u} (p_2) \gamma^\nu \gamma^5 u (p_1)\,,
\end{split}
\end{equation}
where $m_N$, $m_u$, $m_d$ are the masses of the nucleon, and the up and down quarks respectively, $G_{5, N}^q (Q^2)$ is the pseudoscalar form factor associated to the quark $q$ = $u$, $d$, as can be found in \cite{DeMarchi:2024riu}, and $Q^2 \equiv - q^2$, where $q$ is the 4-momentum transfer. For more details about the computation, we refer the reader to the appendix.
We assume the new physics to be G-invariant, so that no contributions arise from second-class currents~\cite{Weinberg:1958ut}.

At the highest energies, the dominant contribution arises from deep inelastic scattering (DIS),
$\chi (k_1) + N(p_1) \to \chi (k_2) + X(p_2)$.
Including collinear factorization, the differential cross section takes the form~\cite{DeMarchi:2025uoo}
\begin{equation}
\begin{split}
\frac{d^2\sigma}{dx dy} = & \frac{y}{16 \pi} \frac{Q^2 \sum_q f_q (x, Q^2) \left| \overline{\mathcal{M}_{\chi q}^\mathrm{DIS}} \right|^2}{\left( Q^2 \right)^2 - 4 m_N^2 m_\chi^2 x^2 y^2}\,,
\end{split}
\end{equation}
where $\mathcal{M}_{\chi q}$ is the amplitude for the interaction at the quark level. The mean squared amplitude is,
\begin{equation}
\begin{split}
\left| \overline{\mathcal{M}_{\chi q}^\mathrm{DIS}} \right|^2 = 2 g_D^2 g_{q a}^2 \frac{ \left(Q^2 \right)^2}{y \, \left(Q^2 + m_a^2 \right)^2} \, \left(Q^2 + 2 m_\chi^2 \right) \left( \frac{\left(Q^2 \right)^2}{2 E_\chi^2} + Q^2 \left( 3y - 1 \right) \right)\,,
\end{split}
\end{equation}
where $m_a$ is the mass of the ALP.

In the examples shown in what follows, we will assume for simplicity a universal coupling to quarks, so that $g_{a q} = g_{a N}$ for all $q$.
We are also parametrizing this model phase space in terms of the non-relativistic cross section of dark matter and protons as computed in standard direct detection experiments, $\sigma^\mathrm{DD}_{\chi p}$. The non-relativistic cross section for interactions with neutrons is roughly the same. More details about the computation of this quantity are given in \Cref{sec:DD_cs}. This would help in locating the sensitivity of these heating mechanisms in terms of how standard direct detection experiments present their results.

\subsection{Boosted dark matter flux from blazars}

\begin{figure}
    \centering
    \includegraphics[width=0.49\linewidth]{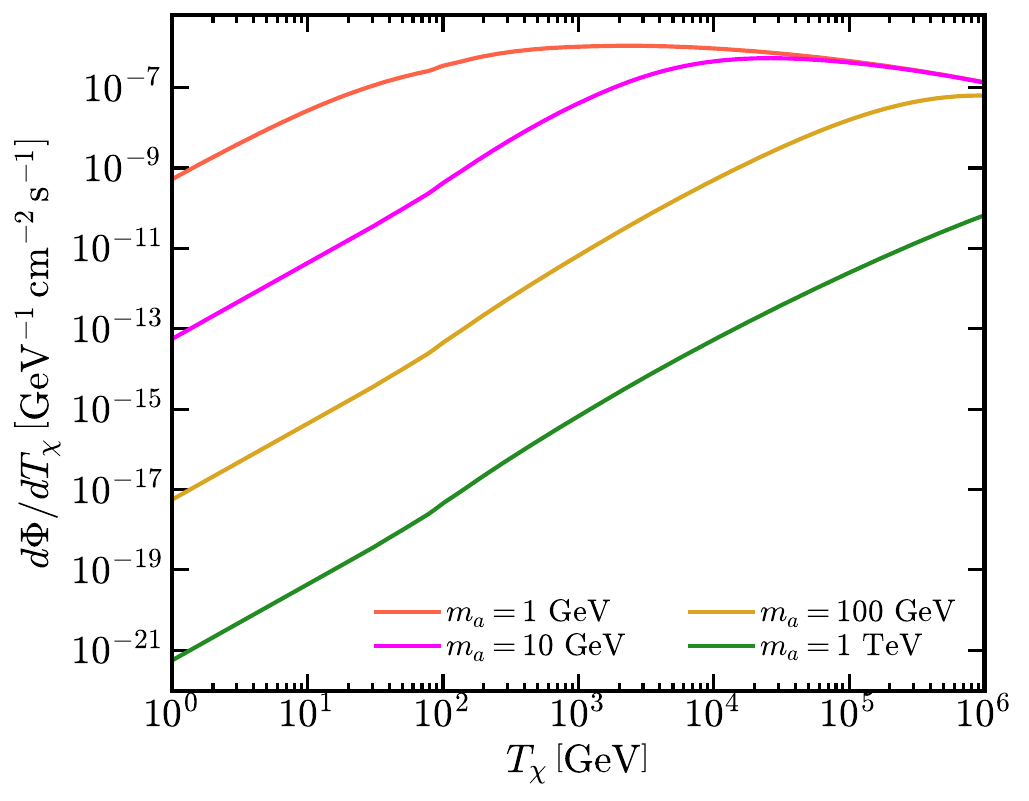}
    \includegraphics[width=0.49\linewidth]{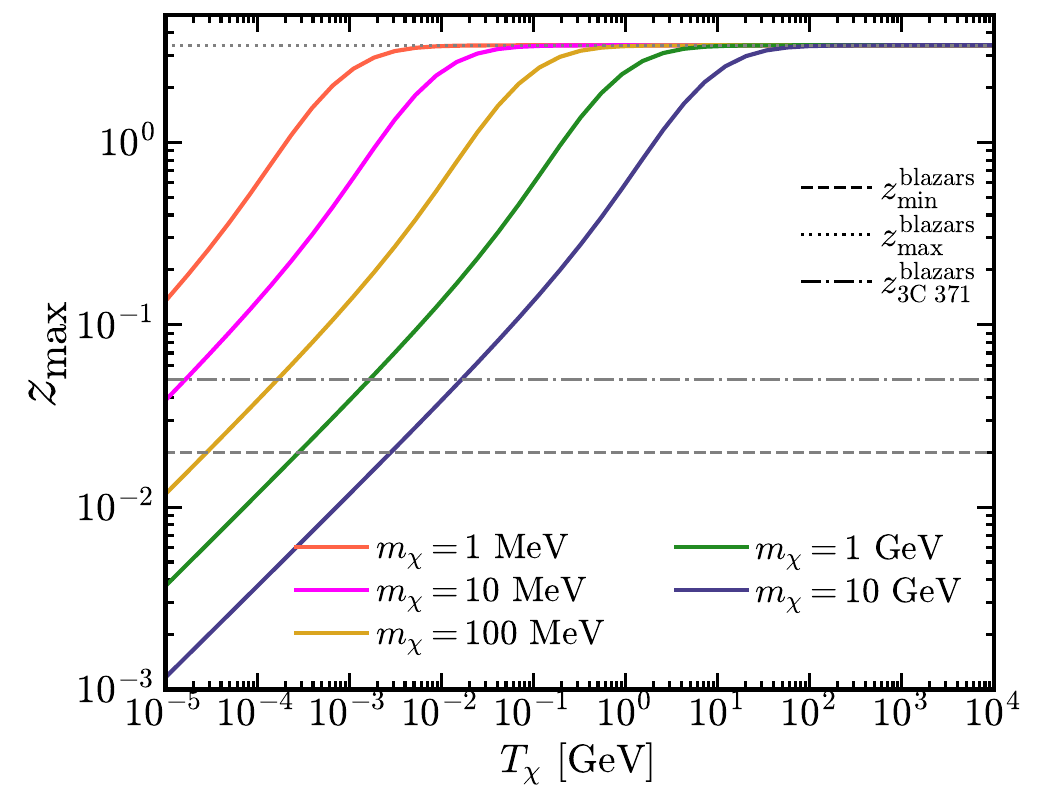}
    \caption{
    Left panel: Fluxes for the ALP with $m_\chi =10\,\mathrm{MeV}$ and
    $m_{a} = 1 \,, 10 \,, 100 \,, 1000$ GeV, after excluding contributions from sources whose DM particles would not reach the Galaxy
    within the blazar lifetime.
    Right panel: Maximum redshift $z$ from which DM particles of different masses and kinetic energies can arrive
    within a typical blazar lifetime ($\sim 10\,\mathrm{Gyr}$). The example of the present work corresponds to the line in magenta, for $m_\chi =10\,\mathrm{MeV}$.
    The dashed gray line indicates the minimum redshift of the 324 blazars considered, the dotted line the maximum,
    and the dash-dotted line highlights the blazar providing the dominant contribution.
    }
    \label{fig:AGNs_flux}
\end{figure}
We consider a BBDM flux originating from a catalogue of 324 blazars~\cite{Rodrigues:2023vbv}.
Blazars produce jets of high-energy protons that can scatter off ambient DM; we neglect contributions from accelerated
electrons.
The differential BBDM flux is given by~\cite{Wang:2021jic, DeMarchi:2025uoo}
\begin{align}
\frac{d\Phi_\chi}{dT_\chi}
=\frac{\Sigma_{\rm DM}}{2\pi m_\chi d_L^2}
\int_0^{2\pi}d\phi 
\int_{T_p^{\rm min}(T_\chi)}^{T_p^{\rm max}}dT_p 
\frac{d\Gamma_p}{dT_p d\Omega}
\frac{d\sigma_{\chi p}}{dT_\chi}~,
\end{align}
with notation as defined in \Cref{sec:BDM_flux}. The DM column density $\Sigma_{\rm DM}$ is
\begin{equation}
\Sigma_{\rm DM}=\int_{R_{\rm min}}^{r} \rho_\chi(r')dr'~,
\end{equation}
where we will consider $R_{\rm min} = 100\, r_s$, with $r_s$ the Schwarzschild radius of the blazar's black hole, as one of the benchmarks in \cite{DeMarchi:2025uoo}.
Details of the flux computation are provided in \Cref{sec:BDM_flux}.

\subsection{Cases, simulation and results}

\begin{figure}
    \centering
    \includegraphics[width=0.49\linewidth]{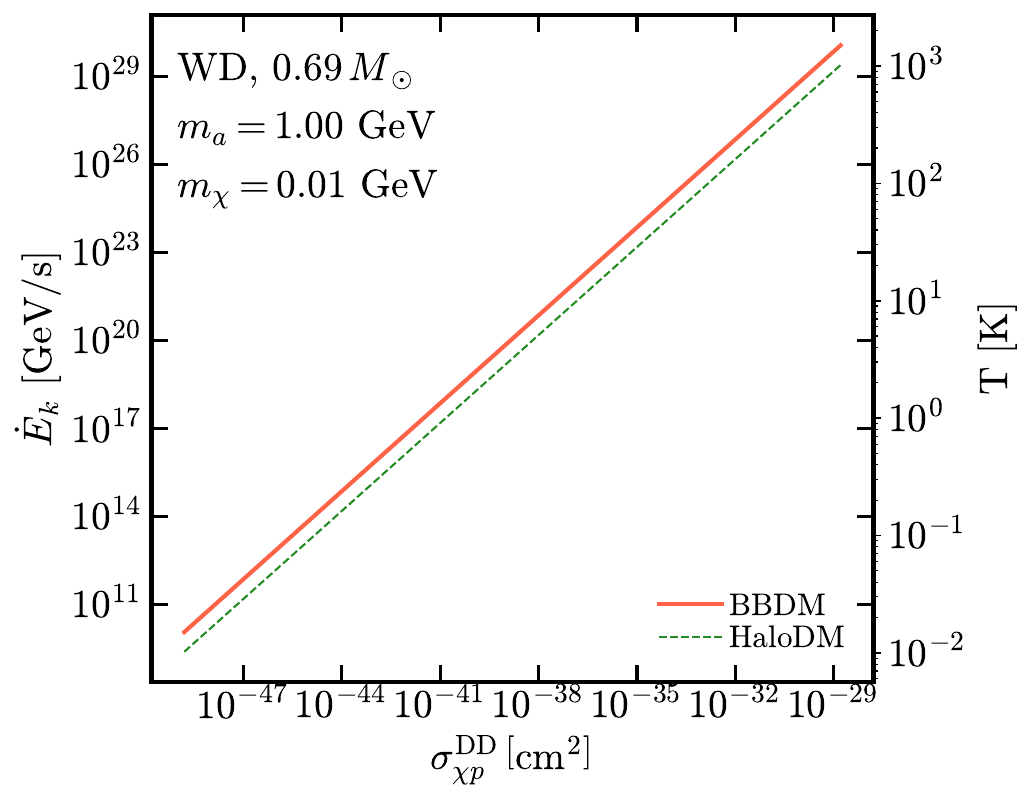}
    \includegraphics[width=0.49\linewidth]{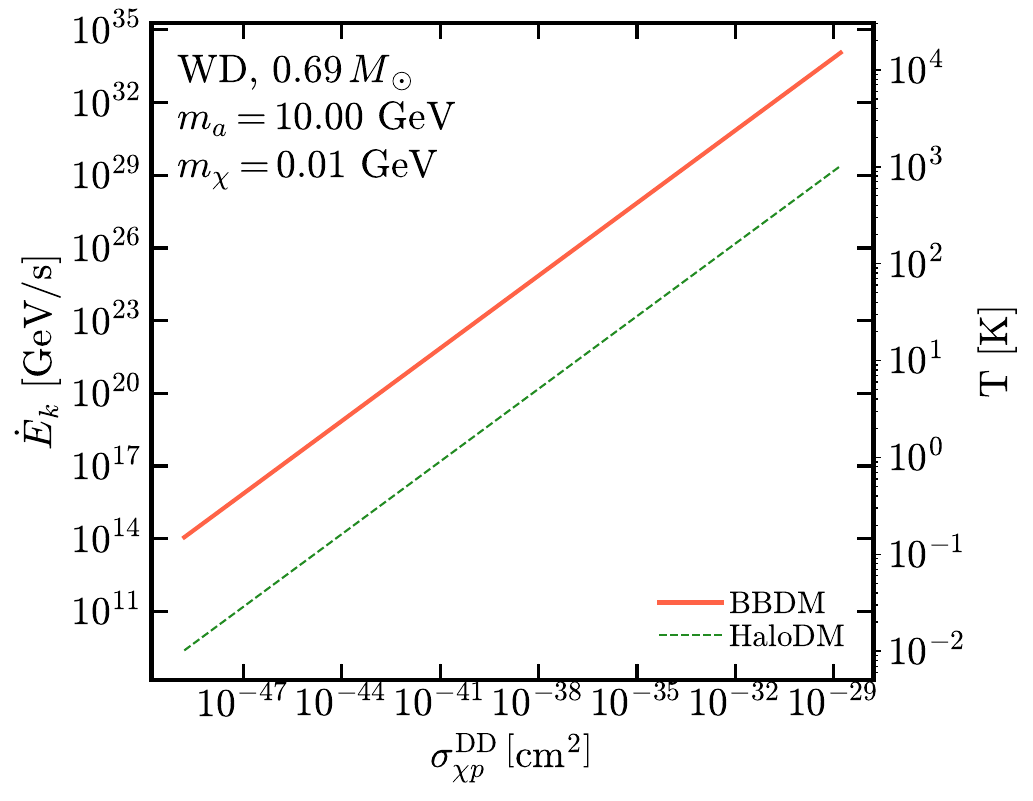}
    \includegraphics[width=0.49\linewidth]{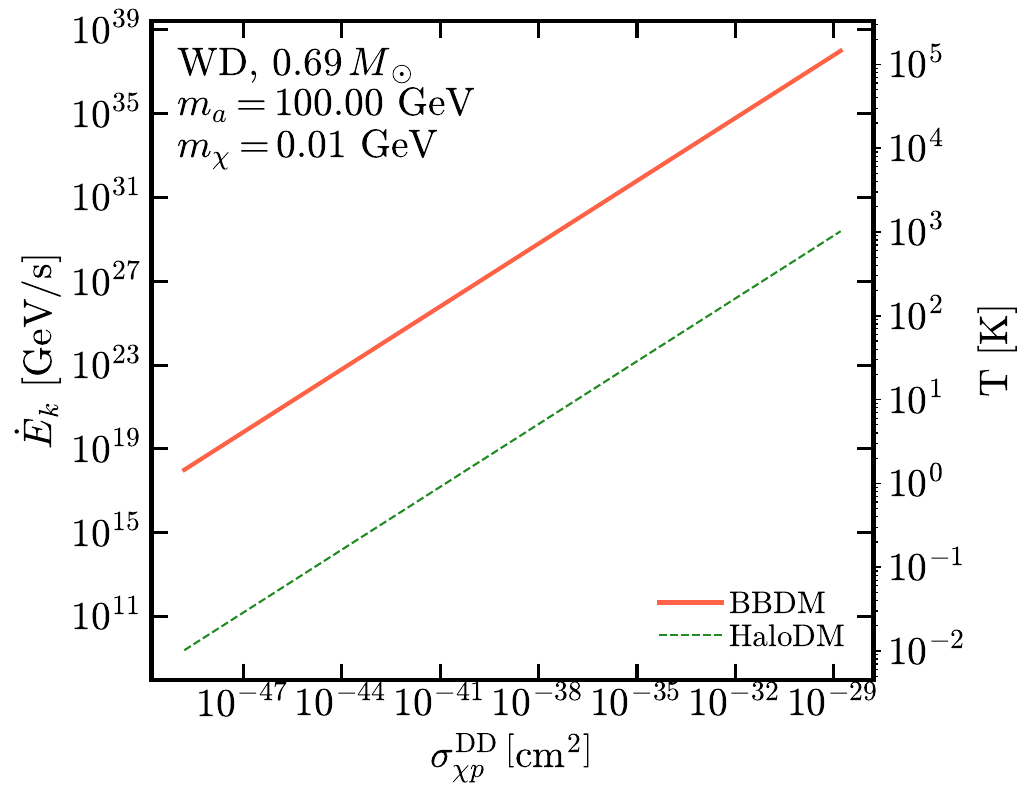}
    \includegraphics[width=0.49\linewidth]{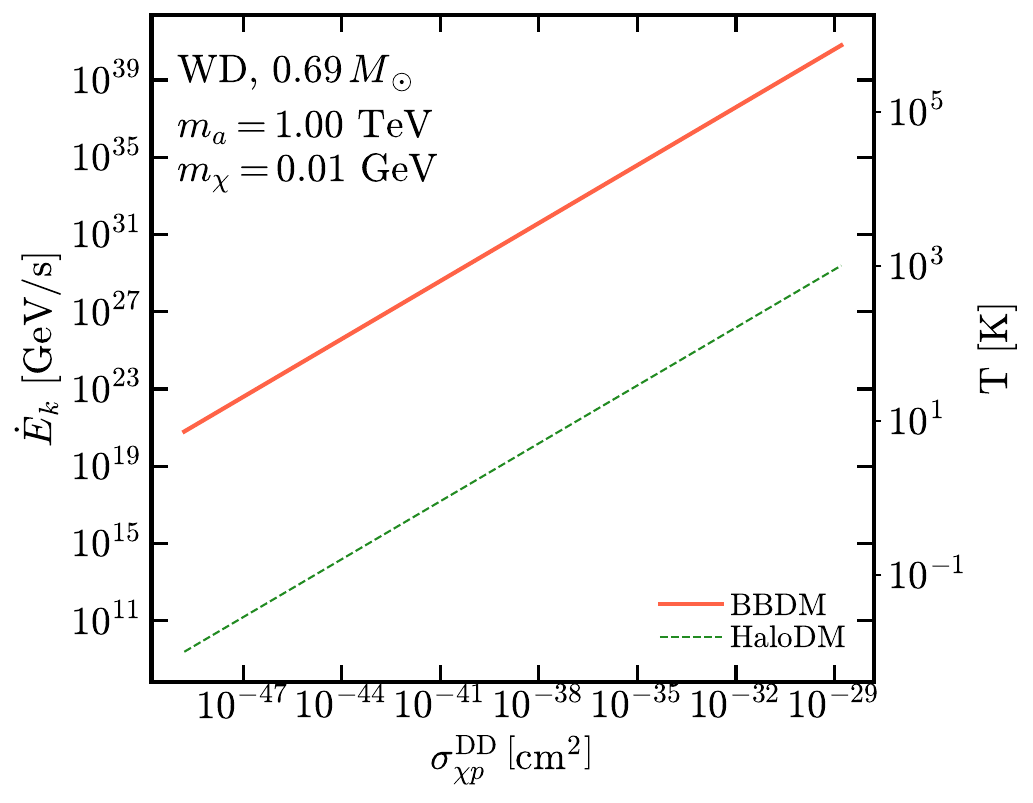}
    \caption{
  Heating of the WD benchmark induced by BBDM,
for $m_\chi = 10~\mathrm{MeV}$ and $m_{a} = 1 \,, 10 \,, 100 \,, 1000$ GeV, for the ALP mediator model.
Dashed lines denote the halo DM contribution.
    }
    \label{fig:Ek_WD}
\end{figure}

We focus on scenarios in the sub-GeV mass range of DM, where direct detection experiments are typically less sensitive.
We choose a benchmark of DM mass of $10$ MeV motivated by the difficulty to constrain those masses from traditional direct detection
experiments and also by the higher number density of DM around blazars that can be boosted for lower mass DM.
We consider four representative cases with that fixed DM mass for different ALP masses: $m_{a} = 1 \,, 10 \,, 100 \,, 1000$ GeV.

The BBDM fluxes associated with these benchmarks are shown in \Cref{fig:AGNs_flux}.
This figure also illustrates the energy cuts applied to the flux, which arise because low-energy DM particles
emitted by distant blazars may not reach the Galaxy within the lifetime of the source.
These cuts depend sensitively on the redshift of each blazar.
We indicate the minimum and maximum redshifts of the catalogue considered~\cite{Rodrigues:2023vbv}, and explicitly highlight
the redshift of the blazar 3C~371, which provides the dominant contribution to the flux.
In general, the flux is larger at lower energies, although it decreases again at very low energies.
DIS processes constitute the dominant contribution to the flux production.

We will mainly focus on white dwarfs, since the energies at which it accelerates dark matter are way below with respect to neutron star energies. Therefore,
halo dark matter capture and heating is more excluded for them. For neutron stars, the cross sections are not strongly suppressed for halo DM heating because they can boost DM up to boost factors of $1.3 - 2$. BBDM heating is not competitive with respect to halo DM heating in NS. We leave the NS example to show the transition from the optically thin regime to the opaque one just as a reference.

\begin{figure}
    \centering
    \includegraphics[width=0.7\linewidth]{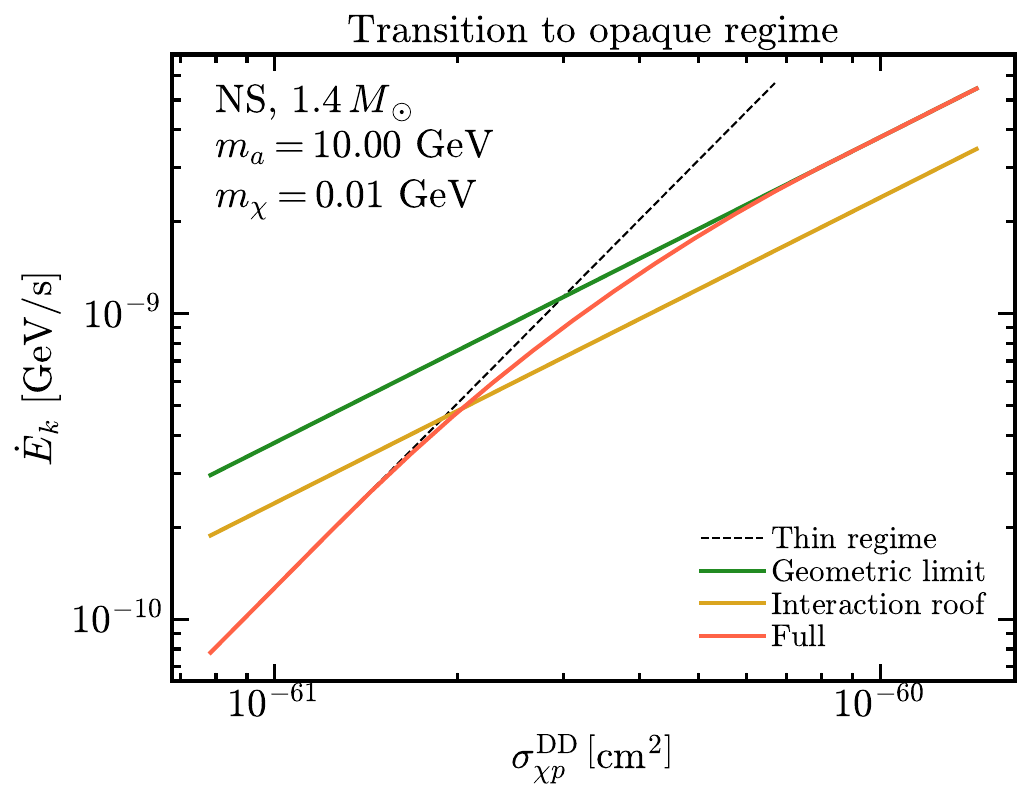}
    \caption{
    Transition of opaque regime in heating of the NS benchmark induced by BBDM,
for $m_\chi = 10~\mathrm{MeV}$ and $m_{a} = 10$ GeV. The parameter space is parametrized in terms of
the standard direct detection non-relativistic cross section, $\sigma_{\chi p}^{\rm DD} \propto g_D^2g_{aN}^2$.
}
    \label{fig:Ek_NS}
\end{figure}
The resulting heating rates for all benchmark scenarios are shown in \Cref{fig:Ek_WD} for white dwarfs. 
In dashed lines we can also see the heating rate from halo DM capture. We can see that BBDM heating can be competitive (for $m_a = 1$ GeV) or be much larger (for $m_a > 1$ GeV)
with respect to the traditional halo DM heating. This is due to the strong suppression of the cross section at low energies for this ALP mediator model.

We consider the plots with respect to non-relativistic cross sections with protons for direct detection experiments, $\sigma^\mathrm{DD}_{\chi p}$,
so that we can have an idea of which regions in direct detection we are exploring with these processes.
For $m_\chi \simeq 1.5\,\mathrm{GeV}$, standard direct detection experiments are sensitive to cross sections above ${\cal O}(10^{-25}\,\mathrm{cm}^2)$, and below $m_\chi \lesssim 1.5\,\mathrm{GeV}$, direct detections cannot give effective constraints~\cite{Cirelli:2013ufw}. In our case, for $m_\chi = 0.01\,\mathrm{GeV}$, the predicted cross section in our scenario lies well below the reach of current direct detection experiment.
The Migdal effect in semiconductors can give constraints to light dark matter. For $m_\chi \approx 0.01~{\rm GeV}$, it can reach $\sigma_{\chi p}^{\rm DD}\approx {\cal O}(10^{-15}~{\rm cm^2})$~\cite{SENSEI:2023zdf,Berghaus:2026kmj}.

In general, at small couplings, the heating rate follows the optically thin behavior, where energy deposition is dominated by
single-scattering events and grows rapidly with the coupling strength. However, for ALP mediated models, the DIS regime
exhibits high cross sections even for low couplings, so that the interactions are saturated in the star.
In the optically thin regime, the heating rate grows as the fourth power of the couplings, $\dot{E}_k \propto g_{D}^4 g_{a N}^4$, since there is a quadratic dependence on the
flux production ($\Phi_\chi \propto g_{D}^2 g_{a N}^2$), and another quadratic one in the interaction rate in the star ($\Omega^- \propto d\sigma/d\Pi \propto g_{D}^2 g_{a N}^2$).
As the coupling increases, the heating departs from the thin-limit scaling and approaches the interaction roof,
signalling the onset of efficient scattering along particle trajectories inside the star. We can see this behavior in \Cref{fig:Ek_NS},
where we needed to plot extremely low cross sections in order to reach the optically thin regime and analyze its
transition to the opaque regime.

For larger couplings, the heating rate saturates at the geometric limit, where essentially all particles
crossing the star deposit all of their energy and the rate becomes independent of the microscopic interaction strength.
There is still a quadratic dependence on the couplings, $g_{D}^2 g_{a N}^2$, since the flux is affected by these quantities.
This saturation occurs at smaller values of the couplings for neutron stars than for white dwarfs, reflecting
the higher densities and more compact geometry of neutron stars. The cross sections are dominated by DIS interactions inside the stars,
and there is no Pauli blocking for these interactions. Therefore, the heating rate efficiency just depends on the strength 
of the cross section and the target number density, values that are higher for neutron stars, due to their compactness and
higher energies of the interactions.

Heavier dark matter masses lead to systematically smaller heating rates,
because the fluxes are suppressed for large DM masses ($\Phi_\chi \propto \rho_\chi / m_\chi$), although the kinetic energy available for deposition is higher for those heavier cases.
This also motivated us to study a light DM case: $10$ MeV.

\begin{figure}
    \centering
    \includegraphics[width=0.7\linewidth]{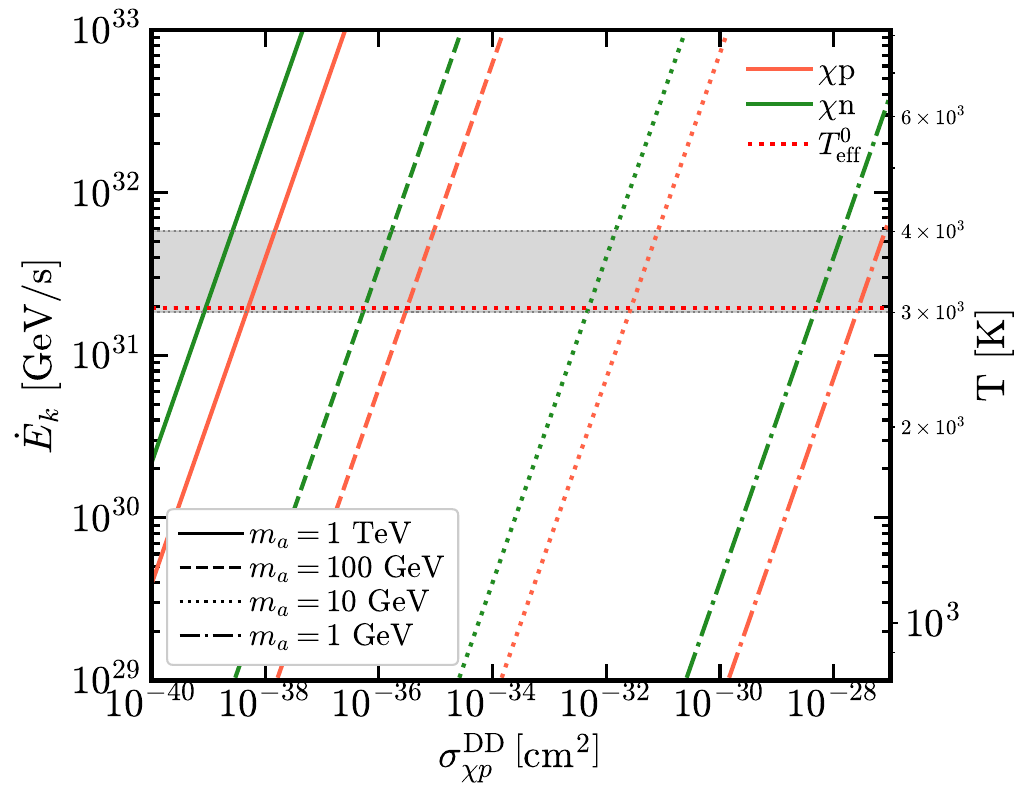}
    \caption{
    Constraints from heating by considering the coolest observed WDs, with temperatures between $3000$ and $4000$ K, for $m_\chi = 10~\mathrm{MeV}$ and $m_{a} = 1, 10, 100$ GeV and $1$ TeV.
    We particularly show in a red dotted line the temperature of the selected WD, $3048$ K, labeled as $T_\mathrm{eff}^0$.
    The parameter space is parametrized in terms of
    the standard direct detection non-relativistic cross section, $\sigma_{\chi N}^{\rm DD}$, with protons ($N = p$) and neutrons ($N = n$).
}
    \label{fig:cs_limits}
\end{figure}

Finally, if we consider the coolest WDs ever observed, with temperatures between $3000$ and $4000$ K by the James Webb Space Telescope (JWST)~\cite{blouin2024jwst}, then we can infer limits around these temperatures, which would represent a region above which would be excluded. This is especially relevant for WD J2147-4035, since it is the coolest ever observed~\cite{Elms:2022oic}. We are mostly sensitive to the heavier ALP-mediator cases. For $m_a = 1$ TeV, we hit those temperatures for $\sigma^\mathrm{DD}_{\chi N} \sim 10^{-39} - 10^{-38} \, \mathrm{cm}^2$; for $m_a = 100$ GeV, for $\sigma^\mathrm{DD}_{\chi N} \sim 10^{-36} - 10^{-35} \, \mathrm{cm}^2$; for $m_a = 10$ GeV, for $\sigma^\mathrm{DD}_{\chi N} \sim 10^{-33} - 10^{-31} \, \mathrm{cm}^2$; for $m_a = 1$ GeV, for $\sigma^\mathrm{DD}_{\chi N} \sim 10^{-29} - 10^{-27} \, \mathrm{cm}^2$. In the particular case of WD J2147-4035, the constraints are: $\sigma^\mathrm{DD}_{\chi p} = 5.0 \times 10^{-39} \, \mathrm{cm}^2$ and $\sigma^\mathrm{DD}_{\chi n} = 9.0 \times 10^{-40} \, \mathrm{cm}^2$ ($m_a = 1$ TeV); $\sigma^\mathrm{DD}_{\chi p} = 3.1 \times 10^{-36} \, \mathrm{cm}^2$ and $\sigma^\mathrm{DD}_{\chi n} = 5.6 \times 10^{-37} \, \mathrm{cm}^2$ ($m_a = 100$ GeV); $\sigma^\mathrm{DD}_{\chi p} = 2.7 \times 10^{-32} \, \mathrm{cm}^2$ and $\sigma^\mathrm{DD}_{\chi n} = 4.9 \times 10^{-33} \, \mathrm{cm}^2$ ($m_a = 10$ GeV); $\sigma^\mathrm{DD}_{\chi p} = 2.7 \times 10^{-28} \, \mathrm{cm}^2$ and $\sigma^\mathrm{DD}_{\chi n} = 4.9 \times 10^{-29} \, \mathrm{cm}^2$ ($m_a = 1$ GeV). These values are much smaller than the standard direct detection limits, although they rely on some assumptions about blazar boosting mechanisms, such as the column density considered and the estimations of proton jets at each AGN.

\section{Conclusions}
\label{sec:conclusions}

In this work, we have developed and applied a general framework to study the heating of compact stars induced by DM jets. As shown throughout the paper, the same formalism can be extended to isotropic fluxes, provided the stellar geometry is spherically symmetric. We constructed a fully relativistic and geometry-aware description of energy deposition by dark matter in compact stars, consistently accounting for geodesic propagation, gravitational focusing, optical-depth effects, and interactions in degenerate matter. Heating is treated as a general energy-deposition process that includes both gravitational capture and scattering by particles that traverse the star without becoming bound, thereby extending beyond capture-only descriptions. While blazar-sourced boosted DM served as our primary motivation, the framework is not restricted to a specific energy range. At sufficiently low energies, however, the treatment of geodesic congruences becomes technically more involved.

Within this framework, we identified three distinct regimes governing the heating of compact stars: the optically thin limit, the interaction roof, and the geometric limit. The transitions between these regimes are controlled by the interplay between microscopic particle-physics couplings\footnote{The couplings may also affect the incoming flux, such as in the examples treated in \Cref{sec:examples}.} and the macroscopic structure of the star. Our results explicitly demonstrate how the full calculation interpolates between these limiting behaviours once optical-depth effects and geodesic amplification are treated consistently. It is also worth mentioning that for very energetic fluxes, the only relevant regime may be the geometric limit, due to the high probability rate of interaction as cross sections grow with energies.

To illustrate these effects in concrete and well-controlled astrophysical settings, we focused on compact stellar remnants, such as white dwarfs and neutron stars. These objects are well described by spherically symmetric solutions of the TOV equations with established equations of state and span a wide range of densities, radii, and gravitational potentials.

As a representative case study, we applied the formalism to blazar-sourced boosted dark matter. In this setup, different interaction mechanisms govern energy deposition at different energies: elastic scattering dominates at low energies, while inelastic interactions control the transfer of energy at sufficiently high energies, driving the approach to the interaction roof and the geometric limit. Within this unified framework, neutron stars reach both saturation regimes at smaller couplings than white dwarfs, reflecting their higher densities and stronger gravitational fields. This contrast highlights the complementary response of different compact objects to dark-matter-induced heating. Due to the higher energies, the geometric limit is reached much more easily than for halo dark matter heating. Since in the geometric limit, white dwarfs can interact with more dark matter particles than neutron stars, due to their much larger volumes, these are favoured scenarios for studying boosted dark matter heating stars, and can constrain dark matter models in a much stronger way than halo dark matter heating.

As a concrete particle-physics realization, we employed fermionic DM interacting with SM quarks through an ALP mediator, as described by the Lagrangian in \Cref{eq:Lag}. The cross sections in this model scale with powers of the transferred momentum, rendering them strongly suppressed at low energies. This feature makes ALP-mediated interactions particularly challenging to constrain with traditional direct detection experiments, especially for sub-GeV DM masses such as the $m_\chi = 10\,\mathrm{MeV}$ benchmark adopted in this work. We showed that, for this class of models, BBDM heating of white dwarfs can be competitive with (for $m_a = 1\,\mathrm{GeV}$) or significantly exceed (for $m_a > 1\,\mathrm{GeV}$) the standard halo DM heating contribution. This is a direct consequence of the boosted population probing the high-energy regime, where the momentum-dependent ALP couplings are no longer suppressed and DIS interactions dominate. In contrast, halo DM interacts at energies where the ALP-mediated cross sections are inefficient, yielding much weaker heating rates.

Heated compact stars, and especially white dwarfs, can be observed by James Webb Space Telescope (JWST)~\cite{Baryakhtar:2017dbj}, which can measure temperatures of $\mathcal{O} (10^3$ K) for white dwarfs~\cite{blouin2024jwst}. This allows us to reach sensitivities of $\sigma^\mathrm{DD}_{\chi N} \sim 10^{-39} - 10^{-38} \, \mathrm{cm}^2$ for $m_a = 1$ TeV,  $\sim 10^{-36} - 10^{-35} \, \mathrm{cm}^2$ for $m_a = 100$ GeV, $\sim 10^{-33} - 10^{-31} \, \mathrm{cm}^2$ for $m_a = 10$ GeV, and $\sim 10^{-29} - 10^{-27} \, \mathrm{cm}^2$ for $m_a = 1$ GeV, which constrain these models much more strongly than with standard direct detection experiments, which have sensitivities of the order of $\sigma_{\chi N}^{\rm DD}\approx {\cal O}(10^{-15}~{\rm cm^2})$~\cite{SENSEI:2023zdf,Berghaus:2026kmj}.

The results presented here rely on a set of controlled approximations, including the separation between single-scattering heating and the geometric limit, and the assumption of spherical symmetry. These approximations do not affect the qualitative behaviour identified across the different heating regimes and can be systematically relaxed as improved microphysical treatments and more general geometries are incorporated.

The formalism presented here can be extended to other dark-sector models, alternative flux origins, additional classes of astrophysical targets, and more general flux geometries. The specific benchmark scenarios studied in this work yield heating rates that are competitive and much stronger than standard halo dark matter heating, though the framework provides a general and reusable foundation for exploring a wide range of particle-physics models and flux realizations, and for connecting dark-matter-induced heating to future observations of compact stars. This is particularly relevant for models with low energy suppressed cross sections.

\acknowledgments
We kindly thank Prof. Przemysław Małkiewicz for useful discussions on gravitational effects on volumes and geodesics.
SH would also like to thank Dr. Hailin Xu for useful discussions on DM direct detections.
This research was supported by the Cluster of Excellence ``Precision Physics, Fundamental Interactions, and Structure of Matter'' (PRISMA$^+$, EXC 2118/1), funded by the Deutsche Forschungsgemeinschaft (DFG, German Research Foundation) under the German Excellence Strategy (Project No. 390831469, MRQ),
the National Science Centre, Poland (Grant No. 2021/42/E/ST2/00031, JHZ),
and by support from Prof. Jason L. Evans and Prof. Yuichiro Nakai, funded by the National Natural Science Foundation of China (SH).

\appendix
%
\section{Blazar Boosted Dark Matter–Induced Flux}
\label{sec:BDM_flux}

Active galactic Nuclei (AGNs) are among the most luminous persistent sources in the Universe, powered by accretion of matter onto supermassive black holes. Their high-energy emission spans from radio to gamma rays and has long being recognized as a potential probe of particle physics beyond the SM. 
When an AGN's relativistic jet is pointed close to our line of sight, the object is classified as a blazar.
In particular, blazars can act as powerful accelerators of charged particles, and thus many provide sizeable fluxes of blazar boosted DM (BBDM).

For this work, we rely on a well-defined population of nearby blazars. The sample originates from the CGRABS\footnote{Catalogue can be found in \url{https://heasarc.gsfc.nasa.gov/W3Browse/radio-catalog/cgrabs.html}} catalogue\cite{Healey:2007gb}, which provides a uniform all-sky sample of 1625 blazar candidates with flat radio spectra. From this catalogue, a subset of 324 blazars has been extensively studied in recent works~\cite{DeMarchi:2025uoo,Rodrigues:2023vbv}. This reduced sample represents the brightest gamma-ray blazars with well-measured redshifts and spectral information, and has become a standard dataset in multimessenger analyses.

When  blazars are considered as sources of boosted projectiles, the particle emission along the jet axis can be described as~\cite{Wang:2021jic,DeMarchi:2025uoo}
\begin{equation}
    \frac{d\Gamma_i}{dT_id\Omega} = \frac{k_i}{4\pi}\biggl( 1+\frac{T_i}{m_i} \biggr)^{-\alpha_i}\frac{\beta_i(1-\beta_i\beta_B\mu)^{-\alpha_i}\Gamma_B^{-\alpha_i}e^{-(m_i+T_i)/(m_i+T_i^{\rm max})}}{\sqrt{(1-\beta_i\beta_B\mu)^2-(1-\beta_i^2)(1-\beta_B^2)}}~,
\end{equation}
where $\Gamma_B=(1-\beta_B^2)^{-1/2}$, $\frac{1}{4\pi}\frac{d\Phi_i}{dT_i}=\frac{1}{2\pi d_L^2}\frac{d\Gamma_i}{dT_id\Omega}\delta(\cos{\theta})$ and $i$ stands for the component of the jet, which may be protons or electrons. The resulting BBDM flux is
\begin{align}
    \frac{d\Phi_\chi}{dT_\chi}
    =\frac{\Sigma_{\rm DM}}{2\pi m_\chi d_L^2}\sum_i \int_0^{2\pi}d\phi 
    \int_{T_i^{\rm min}(T_\chi)}^{T_i^{\rm max}}dT_i 
    \frac{d\Gamma_i}{dT_id\Omega}\frac{d\sigma_{\chi i}}{dT_\chi}~,
\end{align}
with $T_i^{\rm max}=10^8~{\rm GeV}$, if we are just working with protons~\cite{DeMarchi:2025uoo}. The column density is defined as
\begin{equation}
    \Sigma_{\rm DM}=\int_{R_{\rm min}}^{r} \rho_\chi(r')dr'~,
\end{equation}
where we consider benchmark value for $R_{\rm min}=100 \, r_s$ following Refs.~\cite{DeMarchi:2025uoo}.

The computation of the boosted flux requires specifying the DM–SM interaction. As a representative case, we are considering fermionic DM interacting through an ALP mediator, such as described in \Cref{sec:examples}.

In the BBDM flux computation, we include two regimes for $\chi N$ scattering: elastic (EL) and and deep inelastic (DIS). Focusing on the nucleon target $i = N$, the differential cross section w.r.t. $T_\chi$ is organized as
\begin{equation}
\begin{split}
\frac{d\sigma_{\chi N}}{dT_\chi} &= 
\underbrace{\frac{dQ^2}{dT_\chi} \frac{d\sigma_\mathrm{el}}{dQ^2}}_{\rm EL}
\;+\;
\underbrace{\int_{\nu_\mathrm{min}}^{\nu_\mathrm{max}} d \nu \,
\frac{d\sigma^2_\mathrm{DIS}}{dx dy} 
\left| \frac{\partial (x, y)}{\partial (T_\chi, \nu)} \right|}_{\rm DIS},
\end{split}
\end{equation}
where $\nu \equiv q_0$ is the energy transferred in the interaction, $W$ is the invariant mass of the outgoing particle(s) in the hadronic current, and $Q^2\equiv 2 m_\chi T_\chi$.

\paragraph*{DIS mapping.} Projecting the $\nu$–integral onto the DIS $(x,y)$ plane yields a hyperbola because $Q^2$ is fixed and $x y = Q^2 / (2 E_\chi m_N)$ is constant. The limits $\nu_\mathrm{min/max}$ are:
\begin{equation}
\begin{split}
\nu_\mathrm{min} &= \frac{  m_\chi T_{\chi }}{m_p}\,,\\
\nu_\mathrm{max} &= \min \left( E_{\chi }^{\mathrm{Lab}} - m_\chi, \frac{m_\chi T_p}{m_p} \right)\,.
\end{split}
\end{equation}

In case of computing interactions with nuclei in the star, we neglect spectral functions and use the impulse approximation
\begin{equation}
\frac{d\sigma_{\chi \mathrm{(A,Z)}}}{d\Pi} \;=\; Z \,\frac{d\sigma_{\chi p}}{d\Pi} + (A - Z)\, \frac{d\sigma_{\chi n}}{d\Pi}\,,
\end{equation}
where $\Pi$ represents the parameter space variables of the cross section involved. This approach suffices away from very low energies.

As summarized in Tables~\ref{tab:agn-summary} and \ref{tab:agn-top}, 
our working sample consists of 324 blazars with measured redshifts and 
luminosity distances, among which a handful of nearby blazars dominate 
the stacked contribution. To connect the microscopic flux definition with astrophysical sources, we adopt a flat $\Lambda$CDM background with $\Omega_m=0.315$, $\Omega_\Lambda=0.685$, and $H_0=70.2~\mathrm{km\,s^{-1}\,Mpc^{-1}}$. For each blazar at redshift $z_j$ we compute the propagation time of a boosted particle with mass $m_\chi$ and present-day kinetic energy $T_{\chi,0}$,
\begin{equation}
t_\chi(z, m_\chi, T_\chi) = \int_0^{z} \frac{dz'}{(1+z') H(z')}
\sqrt{1+\frac{m_\chi^2}{T_\chi(z)[T_\chi(z)+2m_\chi]}\left(\frac{1+z}{1+z'}\right)^2},
\end{equation}
with $T_\chi(z)$ obtained from the usual redshifting of momentum and energy.
We then estimate the age of the blazar by computing the time light needs to get to us by considering the blazar's redshift
and whenever this value is greater than $10$ Gyr, we set it as the age of the blazar. If not, we take $10$ Gyr as the standard age
of blazars. Therefore, we have
\begin{equation}
t_\mathrm{BH}^i = \max\big( 10 \, \mathrm{Gyr},\;t_\gamma(z_i)\big),
\end{equation}
where $t_\gamma(z_i)$ is the time light takes to travel to us, 
\begin{equation}
t_\gamma(z) = \int_0^{z} \frac{dz'}{(1+z') H(z')}.
\end{equation}
We impose the condition $t_\mathrm{BH}^i > t_\chi(z_i, m_\chi, T_\chi)$, ensuring that only sources from which particles of mass $m_\chi$ and kinetic energy $T_\chi$ have enough time to reach our galaxy contribute to the stacked flux.

The total flux is then
\begin{equation}
\frac{d\Phi_\chi}{dT_\chi}(T_\chi;m_\chi) \;=\;
\sum_{j=1}^N
\Theta\!\big(t_\mathrm{BH}^j - t_\chi(z_j, m_\chi, T_\chi) \big)
\left(
\left.\frac{d\Phi_{\mathrm{EL}}}{dT_\chi}\right|_j
+
\left.\frac{d\Phi_{\mathrm{DIS}}}{dT_\chi}\right|_j
\right).
\label{eq:CRDM_flux_Tchi}
\end{equation}

Given the typical age of blazars ($\sim 10\,\mathrm{Gyr}$), the right panel of \Cref{fig:AGNs_flux} shows the maximum redshift $z$ from which DM particles of different masses and kinetic energies $T_\chi$ (measured in our Galaxy, not at the source) can contribute to the flux. The dashed and dotted gray lines indicate, respectively, the minimum and maximum redshifts of the 324 blazars in our sample, while the dash-dotted line corresponds to the single blazar that provides the dominant contribution.
These limits follow from the quantities listed in \Cref{tab:agn-summary}, and the ranking proxy $k_p/(4\pi d_L^2)$ is used in \Cref{tab:agn-top} to identify the most relevant sources. 
The left panel of \Cref{fig:AGNs_flux} shows the fluxes for a fixed $m_\chi = 10\,\mathrm{MeV}$ DM mass and four different ALP masses: $m_a = 1, \, 10, \, 100, \, 1000$ GeV, after excluding contributions from blazars whose DM particles could not reach our Galaxy within their lifetime.

Finally, the kinetic energy of the DM fermion, $T_\chi$, is redshifted as the particles reach our galaxy, to a value $T_\chi^{(\infty)}$~\cite{DeMarchi:2025uoo}, where the superscript $(\infty)$ means that we are measuring the energy in the galaxy, but sufficiently far from the star:
\begin{equation}
\begin{split}
T_\chi^{(\infty)} = m_\chi \left( \sqrt{\frac{T_\chi \left( T_\chi + 2m_\chi \right)}{m_\chi^2 \left(1 + z \right)^2} + 1 } - 1 \right)\,,
\end{split}
\end{equation}
where $z$ is the redshift of the blazar. The flux also needs to be expressed in terms of $T_\chi^{(\infty)}$. Therefore, we obtain it as follows:
\begin{equation}
\begin{split}
\frac{d \Phi_{\chi}}{d T_{\chi}^{(\infty)}} = \frac{d \Phi_{\chi}}{d T_{\chi}} \times \frac{d T_{\chi}}{T_{\chi}^{(\infty)}}\,,
\end{split}
\end{equation}
where as in \cite{DeMarchi:2025uoo}: 
\begin{equation}
\frac{d T_{\chi}}{d T_{\chi}^{(\infty)}} =  \frac{\left(1 + z \right)^2 \left( m_\chi + T_{\chi}^{(\infty)}\right)}{\sqrt{ m_\chi^2 + \left(1 + z \right)^2 T_{\chi}^{(\infty)} \left(2 m_\chi + T_{\chi}^{(\infty)} \right)}}\,.
\end{equation}
 \begin{table}[t]
\centering
\caption{Summary of the blazar working sample used in this work. Distances are luminosity distances $d_L$.}
\label{tab:agn-summary}
\begin{tabular}{ll}
\hline
Quantity & Value \\
\hline
Number of sources & 324 \\
$z$ (min, median, max) & (0.02, 1.02, 3.41) \\
$d_L$ [Mpc] (min, median, max) & (85.9, 4.83e+03, 1.47e+04) \\
Ranking proxy  & $k_p/(4\pi d_L^2)$ \\
\hline
\end{tabular}
\end{table}

Table~\ref{tab:agn-summary} summarizes the properties of the blazar working sample, listing the number of sources and the ranges in redshift and luminosity distance. Table~\ref{tab:agn-top} lists the five sources that dominate the stack according to the proxy $k_p/(4\pi d_L^2)$. Nearby TeV blazars (e.g., Mkn~501, PKS~2155$-$304) rise to the top owing to small $d_L$ and large $k_p$, while large $\Gamma_{\rm B}$ and small $\theta_{\rm LOS}$ further enhance their weight. The percentage column shows each source's share of the total proxy stack and highlights which objects control our sensitivity.
\begin{table*}[t]
\centering
\caption{Top contributors according to the proxy $\phi_i \propto k_p/(4\pi d_L^2)$. Values use the provided $k_p$ and $d_L$. However, after the boosting capacity and the line-of-sight are also taken into account, the blazar 3C 371 is the top contributor to the fluxes.}
\label{tab:agn-top}
\resizebox{\textwidth}{!}{
\begin{tabular}{l l c c c c c c c}
\hline
Rank & Source & $z$ & $d_L$ [Mpc] & $\Gamma_{\rm B}$ & $\theta_{\rm LOS}$ [deg] & $L_p$ [erg/s] & $k_p$ [s$^{-1}$ sr$^{-1}$] & \% of stack \\
\hline
1 & Mkn 501 & 0.03 & 129 & 5.8 & 9.88 & 5.01e+46 & 9.91e+41 & 32.5 \\
2 & PKS 2155-304 & 0.12 & 527 & 4 & 14.3 & 1.58e+47 & 6.59e+42 & 13 \\
3 & TXS 0214+083 & 0.09 & 393 & 7.4 & 7.74 & 3.16e+47 & 3.05e+42 & 10.8 \\
4 & 3C 371 & 0.05 & 216 & 9.2 & 6.23 & 1e+47 & 7.86e+41 & 9.19 \\
5 & Mkn 180 & 0.04 & 173 & 6.2 & 9.24 & 2e+46 & 3.45e+41 & 6.34 \\
\hline

\end{tabular}}
\end{table*}

The distribution of these blazars across the sky, together with their high bolometric luminosities, makes them an ideal population to investigate their cumulative contribution to diffuse signals, as well as the dominant role of individual nearby sources. In particular, stacking analysis over this sample have been used both in multimessenger neutrino studies and in exploring the possible flux of BBDM.

\section{Non-relativistic elastic cross sections with nucleons}
\label{sec:DD_cs}

In order to compute the cross sections that are relevant for direct detection experiments, we need to go from the level of interactions with quarks, to that of interactions with nucleons and in a non-relativistic regime. At low energies, when $Q^2 \ll m_a^2$:
\begin{equation}
\begin{split}
\mathcal{L}^q_\mathrm{eff} &= - \frac{g_D}{m_a^2} \sum_q g_{q a} \left[\overline{\chi} \gamma^\mu \gamma_5 \chi \right] \left[ \overline{q} \gamma^\nu \gamma_5 q \right] q_\mu q_\nu\,.
\end{split}
\end{equation}

We can further simplify this expression by considering that at low energies: $\left[ \overline{P} \gamma^\mu \gamma_5 P \right] q_\mu \to - 2 i m_P \left[ \overline{P} \gamma_5 P \right]$, where $P$ is any fermion.\footnote{See for instance Equation $1$ in \cite{Bollig:2020xdr}.} Then,
\begin{equation}
\begin{split}
\mathcal{L}^q_\mathrm{eff} &\simeq - \frac{4 \, m_\chi \, g_D}{m_a^2} \sum_q m_q \, g_{q a} \left[\overline{\chi} \gamma_5 \chi \right] \left[ \overline{q} \gamma_5 q \right]\,.
\end{split}
\end{equation}

This Lagrangian can be expressed in terms of an interaction with nucleons:
\begin{equation}
\begin{split}
\mathcal{L}^N_\mathrm{eff} &= - \frac{4 \, m_\chi \, g_D}{m_a^2}  c_N \left[\overline{\chi} \gamma_5 \chi \right] \left[ \overline{N} \gamma_5 N \right]\,,
\end{split}
\end{equation}
where $c_N$ is~\cite{Cirelli:2013ufw}:
\begin{equation}
\begin{split}
c_N = \sum_q \frac{m_N}{m_q} \left( m_q \, g_{q a} - A_N \right) \, \Delta_q^N \,,
\end{split}
\end{equation}
and $A_N \equiv \left( m_u^{-1} + m_d^{-1} + m_s^{-1} \right)^{-1} \sum_q g_{q a} / m_q$ and $\Delta_q^p = 0.842$, $-0.427$, $-0.085$ for $q = u$, $d$, $s$, respectively, and for the neutron we replace $u \leftrightarrow d$.

To get the non-relativistic operator from this effective interaction Lagrangian, we need to expand the bispinors and neglect quadratic terms of the momenta, such that,
\begin{equation}
\begin{split}
\left[\overline{\chi} \gamma_5 \chi \right] \left[ \overline{N} \gamma_5 N \right] = 4 \left( \vec{s}_\chi \cdot \vec{q} \right) \left( \vec{s}_N \cdot \vec{q} \right) \,,
\end{split}
\end{equation}
where $\vec{s}_\alpha$ is the 3D-spin of $\chi$ or $N$. The operator $\left( \vec{s}_\chi \cdot \vec{q} \right) \left( \vec{s}_N \cdot \vec{q} \right)$ is usually known as $\mathcal{O}_6^\mathrm{NR}$ in the literature~\cite{Fitzpatrick:2012ix, Cirelli:2013ufw, Catena:2015uha}. In the case assumed in the present work, where we fix all of the quark couplings, $g_{q a} \equiv g_{N a}$, then the final non-relativistic interaction Lagrangian is:
\begin{equation}
\begin{split}
\mathcal{L}^\mathrm{NR}_\mathrm{eff} &= - \frac{16 m_\chi \, m_N \, g_D \, g_{N a} \, G_{q N}}{m_a^2} \times \mathcal{O}_6^\mathrm{NR}\,.
\end{split}
\end{equation}
where $G_{q N} \equiv \Delta^N_u + \Delta^N_d + \Delta^N_s - 3\, \left( m_u^{-1} + m_d^{-1} + m_s^{-1} \right)^{-1} \, \left( \Delta^N_u / m_u + \Delta^N_d / m_d + \Delta^N_s / m_s \right)$.

The non-relativistic cross section of DM particles with a nucleon $N$ is~\cite{Lebedev:2014bba, Borschensky:2020olr},
\begin{equation}
\begin{split}
\sigma^\mathrm{DD}_{\chi N} = \frac{16 \left( g_D \, g_{N a} \, G_{q N} \right)^2}{\pi} \times \frac{m_\chi^4 \, m_N^4}{\left( m_\chi + m_N \right)^2 \, m_a^4}\,.
\end{split}
\end{equation}
%




\bibliographystyle{JHEP}
\bibliography{lib}

\end{document}